\documentclass[prb,twocolumn,showpacs,preprintnumbers,amsmath,amssymb]{revtex4}

\usepackage{graphicx}
\usepackage{dcolumn}
\usepackage{caption2}
\usepackage{bm}

\begin{document}

\preprint{APS/123-QED}

\title{Fidelity, fidelity susceptibility and von Neumann entropy to characterize the phase diagram of an
extended Harper model}
\author{Longyan Gong  $^{1,2,3}$}
\thanks{Email address:lygong@njupt.edu.cn.}
\author{Peiqing Tong  $^{2,}$ }
\thanks{Corresponding author. Email address:pqtong@njnu.edu.cn.}
\affiliation{
 $^{1}$Department of Mathematics, Nanjing Normal University, Nanjing 210097,
 China\\
 $^{2}$ Department of Physics, Nanjing Normal University, Nanjing 210097,
 China\\
$^{3}$ Center of Microfluidics Optics and Technology, Department
of Mathematics and Physics, Nanjing University of Posts and
Telecommunications, Nanjing 210003, China
}%
\date{today}
\begin{abstract}
For an extended Harper model, the fidelity for two lowest band edge
states corresponding to different model parameters, the fidelity
susceptibility and the von Neumann entropy of the lowest band edge
states, and the spectrum-averaged von Neumann entropy are studied
numerically, respectively. The fidelity is near one when parameters
are in the same phase or same phase boundary; otherwise it is close
to zero. There are drastic changes in fidelity when one parameter is
at phase boundaries. For fidelity susceptibility the finite scaling
analysis performed, the critical exponents $\alpha$, $\beta$, and
$\nu$ depend on system sizes for the metal-metal phase transition,
while not for the metal-insulator phase transition. For both phase
transitions $\nu/\alpha\approx2$. The von Neumann entropy is near
one for the metallic phase, while small for the insulating phase.
There are sharp changes in von Neumann entropy at phase boundaries.
According to the variation of the fidelity, fidelity susceptibility,
and von Neumann entropy with model parameters, the phase diagram,
which including two metallic phases and one insulating phase
separated by three critical lines with one bicritical point, can be
completely characterized, respectively. These numerical results
indicate that the three quantities are suited for revealing all the
critical phenomena in the model.

\end{abstract}
\maketitle

\section{Introduction}
In recent years, tools from the  quantum-information theory
\cite{ni00,bo00}, specifically the ground state
fidelity\cite{za06} and quantum entanglement \cite{os021,os022},
have been widely exploited to characterize quantum phase
transitions(QPTs)\cite{sa99}. For example, in one dimensional XY
and Dicke model, fidelity between two ground states corresponding
to slightly different values of the parameters drastically
decreases at phase transition points \cite{za06}. Subsequently,
similar properties are also found in fermionic \cite{co07,pa08},
bosonic systems \cite{bu07,oe07} and other various spin systems
\cite{ch07,za07}. Very recently, fidelity susceptibility(FS) (the
second derivative of fidelity) is introduced to signal QPTs in
one-dimensional Hubbard models\cite{yo07,ca08,gu08}, the
Lipkin-Meshkov-Glick model\cite{kw07}, the Kitaev honeycomb
model\cite{ya08}, and various spin systems\cite{ya07,tz08,ch08}.
It is found that FS is more curial than fidelity itself for it
does not depend on the slightly difference values of model
parameters. In Refs.\cite{zh071,zh072}, the fidelity between
arbitrary two ground states is studied in one dimensional quantum
Ising model. Singularities are found in fidelity surfaces for QPTs
\cite{zh072}. The main advantage of the fidelity to identify QPTs
is that \cite{ya07}, it need not a priori knowledge of the order
parameter, topology, etc, since the fidelity is a purely
Hilbert-space geometrical quantity.

   At the same time,  quantum entanglement has been extensively
applied in condensed matter
physics.\cite{za02,gu04,la06,go05,go06,go07,bu06} For example,
quantum entanglement measured by the von Neumann entropy has been
studied in the Hubbard model for the dimer case \cite{za02}, in
the extended Hubbard model for different band filling\cite{gu04},
in quantum small-world networks\cite{go06}, and in low-dimensional
semiconductor systems \cite{bu06}.  It is found that the von
Neumann entropy is suitable for analyzing the interplay between
itinerant and localized features \cite{za02}, as well as
characterizing quantum phase transition\cite{gu04,la06} and the
localization-delocalization transition of electron
states\cite{go05,go06,go07}.

\begin{figure}[ht]
\includegraphics[width=2.5in]{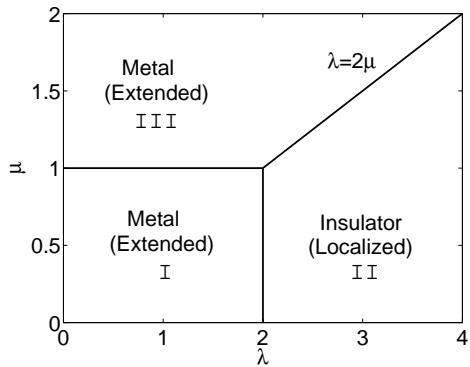}
\caption{Phase diagram of the extended Harper model.}\label{Fig1}
\end{figure}

On the other hand, since the Hofstadter butterfly energy spectrum
was found in 1976\cite{ho76}, the problem of electrons in
two-dimensional periodic potential in a magnetic field has been
attracted much attention\cite{cl79,au80,so85,ha94,ta04,in06} .
After fixing the quasimomentum in one of the directions, a
one-dimensional quasiperiodic system called the Harper model is
deduced\cite{ho76}. The system shows interesting metal-insulator
transitions(MIT)\cite{au80,so85}. Considering the
next-nearest-neighbor hopping for electron on the square lattice
in a uniform magnetic field, an extended Harper model are
proposed\cite{ha94} and studied extensively \cite{ta04}. Very
recently, a similar extended Harper model is introduced from
two-dimensional electrons on the triangular lattice in a uniform
magnetic field\cite{in06}. The phase diagram has a very rich
structure, which shown in Fig.\ref{Fig1}. In regions I and III the
wave functions (spectra) are extended (absolutely continuous), and
in the region II the wave functions are localized (pure points).
On the three boundary lines, the wave functions (spectra) are
critical (singular continuous). Besides the traditional MIT, there
are novel transitions between the two metallic phases(MMT). At the
bicritical point where the triangular lattice symmetry is
retained, both level statistics and multifractal analysis show
quantitively different behaviors from those of other critical
points\cite{in06}.

Considering the above two aspects, we perform a detailed study of
the fidelity between arbitrary two quantum states, FS and von
Neumann entropy for the extended Harper model\cite{in06}. For each
of the three quantities, there are drastic changes at phase
boundaries, i.e., the phase diagram can be distinguished according
to the variations of them with model parameters. Our studies
provide that the two tools, fidelity and von Neumann entropy,
borrowed from the quantum-information theory, are well enough to
identify phase transitions in the system.

The paper is organized as follows. In the next section the
extended Harper model and the definitions of fidelity, FS and von
Neumann entropy are introduced. In Sec.~\ref{sec3} the numerical
results are presented. And we present our conclusions and
discussions in Section ~\ref{sec4}.

\section{\label{sec2}The extended Harper Model, Fidelity, fidelity
susceptibility and von Neumann entropy}

\subsection{\label{sec21} The extended Harper model}
The tight-binding Hamiltonian for an electron moving on a
triangular lattice in a magnetic field \cite{in06} can be reduced
to
\begin{eqnarray}
H= &-&\sum\limits_{n} {[t_a+t_c e^{-2\pi i\phi(n-1/2)+ik_y}]c_n^
\dag
c_{n - 1}}\nonumber\\
&-&\sum\limits_{n} {[t_a+t_c e^{ 2\pi i\phi(n+1/2)-ik_y}]c_n^ \dag
c_{n + 1}}\nonumber\\
&-&2 \sum\limits_{n} {t_b \cos(2\pi\phi n+k_y)c_n^ \dag
c_{n}},\label{form1}
\end{eqnarray} 
where $t_a$, $t_b$ and $t_c$ are the hopping integral for each
bond on the triangular lattice, $\phi$ is a flux that a uniform
magnetic field penetrates each triangle, $k_y$ is a momentum in
the $y$ direction, $c_n^\dag$ ($c_n$) is the
creation(annihilation) operator of the \emph{n}th site in the $x$
direction.

Let $\left|n \right\rangle$ denote $\left| 0,\ldots , 1_{n},
\ldots , 0 \right\rangle$, the general eigenstate of an electron
with eigenenergy $E_\gamma$ is
\begin{equation}
\left| \Psi_{\gamma}  \right\rangle  = \sum\limits_{n }
{\psi^\gamma _n } \left| n \right\rangle  = \sum\limits_{n }
{\psi^\gamma _n c_n^ \dag } \left| 0 \right\rangle,\label{form2}
\end{equation} 
where ${\psi^\gamma_n }$ is the amplitude of the $\gamma$th
eigenstate at the \emph{n}th site. If set $\lambda=
2\frac{t_b}{t_a}$, $\mu=\frac{t_c}{t_a}$ and $t_a$ is taken as
units, the eigenvalue equation  \cite{in06} becomes
\begin{eqnarray}
&-[1+\mu e^{-2\pi i\phi(n-1/2)+ik_y}]\psi_{n-1}
  -[1+\mu e^{ 2\pi i\phi(n+1/2)-ik_y}]\nonumber\\
&\psi_{n+1}-\lambda \cos(2\pi\phi
n+k_y)\psi_{n}=E\psi_{n}.\label{form3}
\end{eqnarray} 
At $\mu=0$ and $\phi$ is irrational, this is reduced to the Harper
equation. Intensively analytical and numerical studies
\cite{au80,so85} for the Harper model show that for $\lambda >2$
the spectrum is pure-point like and all eigenstates are
exponentially localized. For $\lambda <2$ the spectrum becomes
continues with delocalized eigenstates corresponding to ballistic
classical motion. For $\lambda =2$ the situation is critical with
a singular-continuous multifractal spectrum and power law
localized eigenstates. MIT can occur at $\lambda=2$.

\subsection{\label{sec22} Fidelity }
Let $\left| \Psi_{0}(\lambda,\mu) \right\rangle $ denote the
lowest band edge state. According to the definition of
fidelity\cite{ni00,za06,zh071,zh072}, the quantum fidelity (or the
modulus of the overlap of eigenstates) is given by
\begin{eqnarray}
F(\lambda,\mu;\lambda_{0},\mu_{0})&=& |\langle
\Psi_{0}(\lambda,\mu)|\Psi_{0}(\lambda_0,\mu_0)\rangle|.\label{form4}
\end{eqnarray} 
Obviously, $F=1$ if $\lambda=\lambda_{0}$ and $\mu=\mu_0$.

\subsection{\label{sec23} Fidelity susceptibility}
Similarly as that shown in Ref.\cite{za06}, the fidelity for two
lowest edge states with a slightly different parameter values is
defined as
\begin{equation}
F(q)= |\langle \Psi_{0}(q)|\Psi_{0}(q+\delta q)\rangle|.
\label{form5}
\end{equation} 
For simplicity, a certain path $q=q(\lambda,\mu)$ in parameter
spaces can always be supposed. Then the FS can be calculated
as\cite{yo07,ya08,za071}
\begin{equation}
\chi _F  = \mathop {\lim }\limits_{\delta q \to 0} \frac{{ - 2\ln
F(q)}}{{\delta q^2 }}= \sum \limits_{a=\lambda,\mu;b=\lambda,\mu}
g_{a,b}n^a n^b, \label{form6}
\end{equation} 
where $n^\lambda=\partial q/\partial \lambda$ ($n^\mu=\partial
q/\partial \mu$) denotes the tangent units vector at the give
parameter point ($\lambda,\mu$). For the present model,  let
define the driving Hamiltonians
\begin{equation}
 H_\lambda= -\sum\limits_{n} {\cos(2\pi\phi n+k_y)c_n^ \dag
c_{n}}\label{form7}
\end{equation} 
and
\begin{eqnarray}
H_\mu=&&-\sum\limits_{n} {[e^{-2\pi i\phi(n-1/2)+ik_y}]c_n^
\dag c_{n - 1}} \nonumber\\
&&-\sum\limits_{n} {[ e^{ 2\pi i\phi(n+1/2)-ik_y}]c_n^ \dag c_{n +
1}}.  \label{form8}
\end{eqnarray} 
We have
\begin{eqnarray}
g_{ab}  = \sum\limits_{\gamma  \ne 0} {\frac{{\left\langle {\Psi
_\gamma  (q)} \right|H_a \left| {\Psi _0 (q)} \right\rangle
\left\langle {\Psi _0 (q)} \right|H_b \left| {\Psi _\gamma (q)}
\right\rangle }}{{(E_\gamma   - E_0 )^2 }}}.\label{form9}
\end{eqnarray} 

\subsection{\label{sec24} von Neumann entropy}
The general definition of entanglement is based on the von Neumann
entropy \cite{be96}. For an electron in the system, there are two
local states at each site, i.e., $\left| 0 \right\rangle_n, \left|
1 \right\rangle_n$. The local density matrix $\rho_n$ is defined
\cite{za02,gu04,go06,go07} by
\begin{equation}
\rho_n= z_n\left| {1} \right\rangle{_n}{_n}\left\langle {1}
\right| + (1-z_n)\left| {0} \right\rangle{_n}{_n}\left\langle {0}
\right|,\label{form10}
\end{equation} 
where $z_n=\left\langle \Psi_\gamma  \right|c_n^ \dag  c_n \left|
\Psi_\gamma \right\rangle=\left|\psi^\gamma _n \right|^2$ is the
local occupation number at the $n$th site. Consequently, the
corresponding von Neumann entropy related to the \emph{n}th site
is
\begin{equation}
E^\gamma_{vn}=-z_n\log_2z_n-(1-z_n)\log_2(1-z_n).\label{form11}
\end{equation}
For nonuniform systems, the value of $E^\gamma_{vn}$ depends on
the site position $n$. At an eigenstate $\left| \Psi_{\gamma}
\right\rangle$, we define a site-averaged von Neumann entropy
\begin{equation}
E^\gamma_v= \frac{1}{N} \sum\limits_{n=1}^N
{E^\gamma_{vn}},\label{form12}
\end{equation} 
were $N$ is the system size. From the definition (\ref{form12}),
it shows that for an extended state that
$\psi^\gamma_{n}=\frac{1}{\sqrt{N}}$ for all $n$,
 $E^\gamma_v=-\frac{1}{N}\log_2 \frac{1}{N}- (1-\frac{1}{N})\log_2
(1-\frac{1}{N}) \approx \frac{1}{N}\log_2{N}$ at $N
\longrightarrow\infty$, and for a localized state that
$\psi^\gamma_n=\delta_{nn^\circ}$( $n^ \circ$ is a given site ) ,
$E^\gamma_v=0$. In the paper all the values of $E^\gamma_v$ and
$E^\gamma_{vn}$ are scaled by $\frac{1}{N}\log_2{N}$. From the two
examples, we know the scaled $E^\gamma_v$ is near $1$ when
eigenstates are extended, and near zero when eigenstates are
localized. Henceforth, we omit ``scaled'' for simplicity.

In order to analyze the influence of system parameters on the von
Neumann entropy for all the eigenstates, we define a
spectrum-averaged von Neumann entropy as a further gross measure,
i.e.,
\begin{equation}
\langle E_v \rangle  = \frac{1}{M}\sum\limits_{\gamma} {E^\gamma_v
},\label{form13}
\end{equation} 
where $M$ is the number of all the eigenstates.

\section{\label{sec3} numerical results}
In numerical calculations, without loss of generality, we set
$k_y=0$. As a typical case, $\phi=(\sqrt{5}-1)/2$. In fact as is
customary in the context of quasiperiodic system, the value of
$\phi$ may be approximated by the ratio of successive Fibonacci
numbers---$F_m=F_{m-2}+F_{m-1}$ with $F_0=F_1=1$. In this way,
choosing $\phi=F_{m-1}/F_m$ and system size $N=F_m$, we can obtain
the periodic approximant for the quasiperiodic potential. We
directly diagonalize the eigenvalue Eq.(\ref{form2}) at different
values ($\lambda,\mu$) and get all the eigenvalues $E_\gamma$ and
the corresponding eigenstates $\left| \Psi_{\gamma}
\right\rangle$. From the formulas (\ref{form4}---\ref{form13}), we
can obtain the fidelity $F(\lambda,\mu;\lambda_{0},\mu_{0})$, the
FS $\chi_F$, the site-averaged von Neumann entropy $E^{\gamma}_v$
and the spectrum-averaged von Neumann entropy $\langle E_v
\rangle$, respectively. Henceforth, for simplicity we denote $F$
to $F(\lambda,\mu;\lambda_{0},\mu_{0})$ . In all the figures the
system sizes $N$ is chosen to Fibonacci number $987$ unless
specially stated.

\subsection{\label{sec31} Fidelity}


\begin{figure}
\includegraphics[width=2.5in]{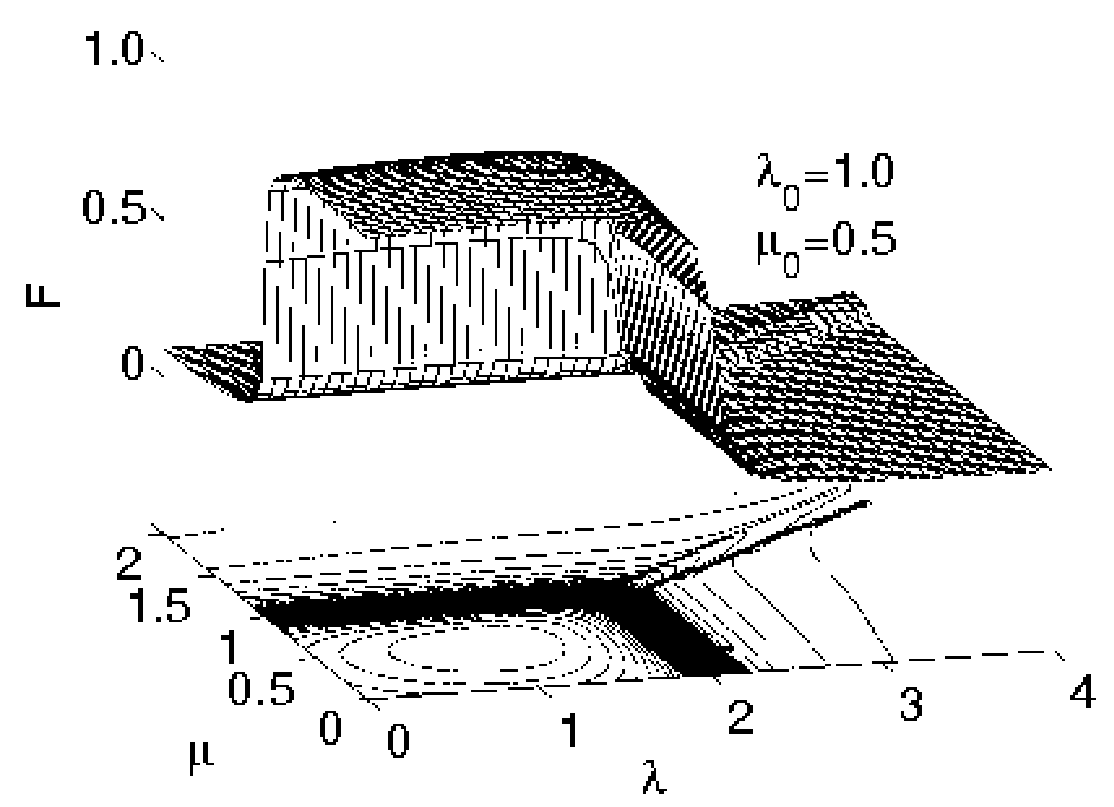}(a)
\includegraphics[width=2.5in]{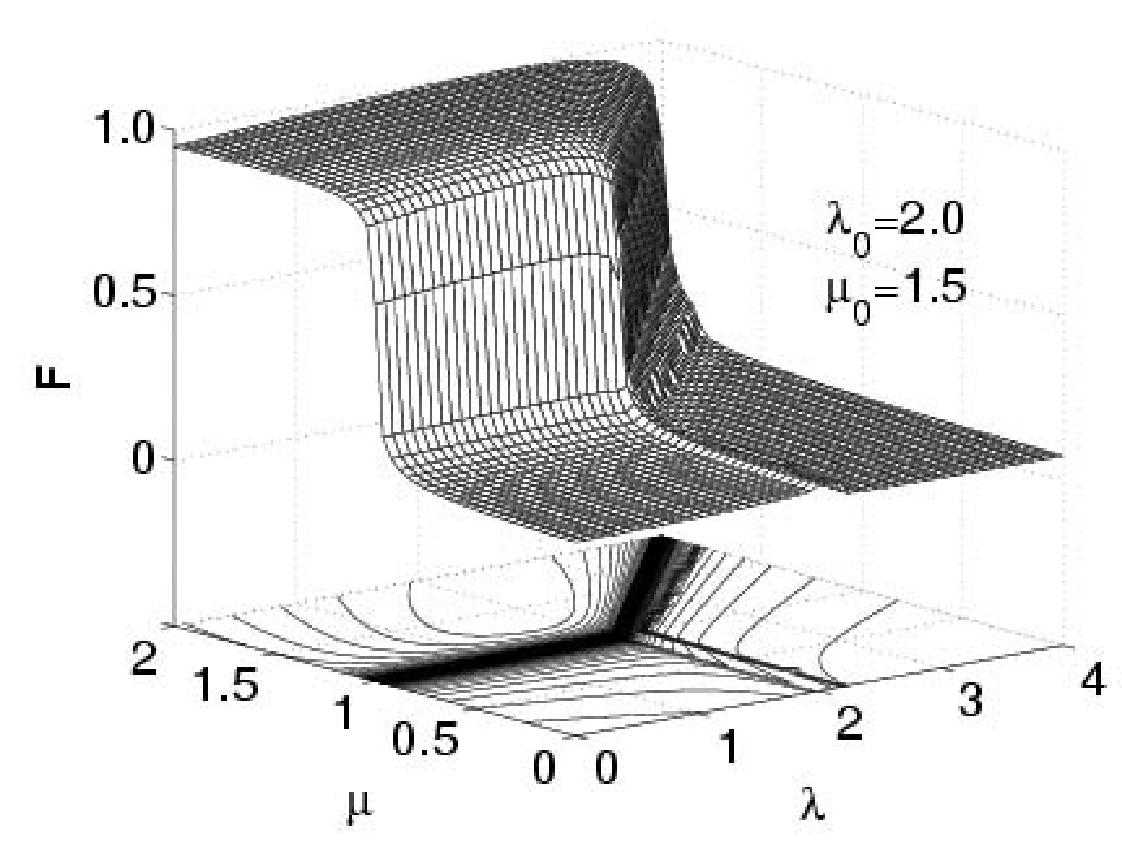}(b)
\includegraphics[width=2.5in]{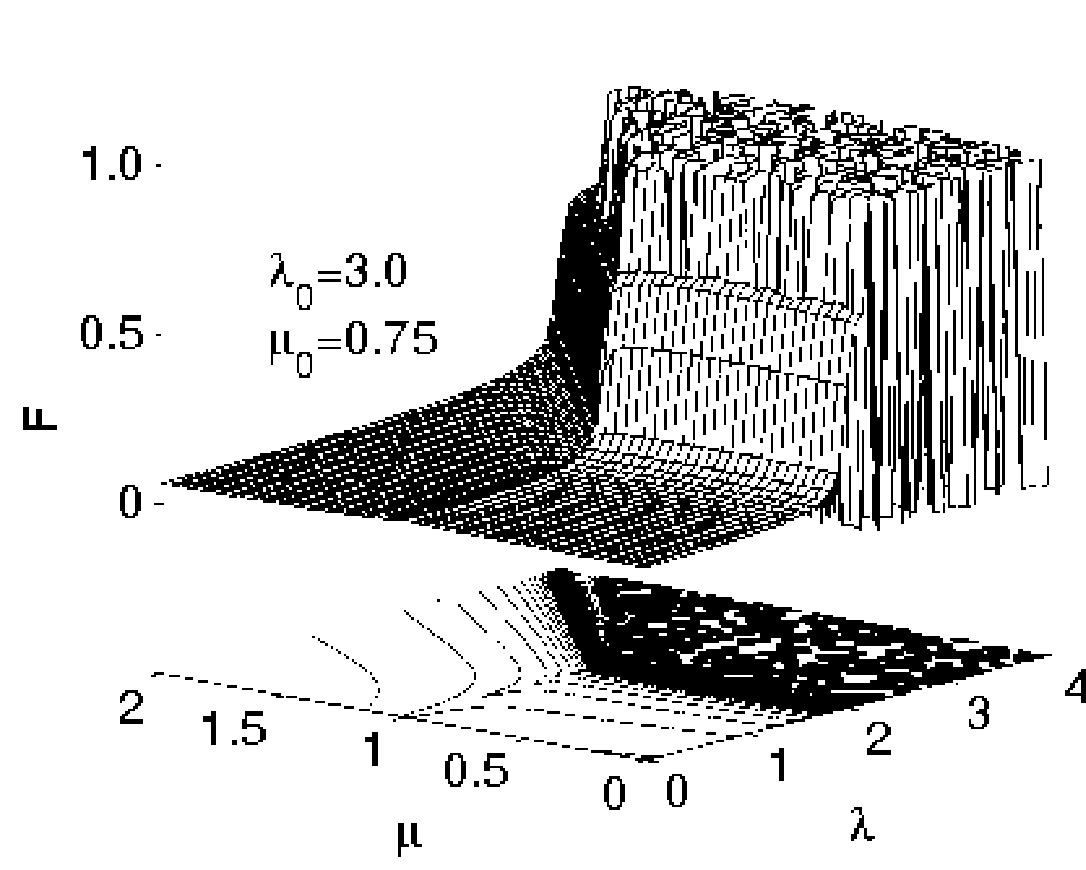}(c)
\caption{The fidelity $F(\lambda,\mu;\lambda_{0},\mu_{0})$ and its
contour map as functions of ($\lambda$,$\mu$) at (a), (b) and (c)
for ($\lambda_0,\mu_0$)=($1.0,0.5$), ($2.0,1.5$) and ($3.0,0.75$),
which corresponding to the system in the metal phase I, the metal
phase III, and the insulator phase II, respectively. }\label{Fig2}
\end{figure}

In the metallic phase I, metallic phase III, and insulating phase
II, we choose ($\lambda_0,\mu_0$)=($1.0,0.5$), ($2.0,1.5$) and
($3.0,0.75$) as examples, respectively. The corresponding fidelity
$F$ varying with parameters $\lambda$ and $\mu$ are shown in
Fig.\ref{Fig2}. At the same time, the contour maps of the fidelity
are also shown. It shows that when parameters are at the same
phase, the fidelity is near one; otherwise, the fidelity is very
small. It is interesting that, though the  phase I and III are
both metallic phases and the corresponding wave functions are all
extended, the fidelity is small when parameters are in the two
phases respectively. This can be understood from the corresponding
``classical orbits" Hamiltonian\cite{in06}: for Phase I, the
contour lines of the Hamiltonian are extended in $x$ direction but
localization in the $y$ direction, while for Phase III, the
contour lines are extended in $x+y$ direction but localized in the
$x-y$ direction. Therefore, the two phases are different. At the
same time, these contour maps divide the parameter space to
different regions, which is good agreement with the phase diagram
shown in Fig.\ref{Fig1}. Comparing with the fidelity shown in
Fig.\ref{Fig2}(a) and (b), the fidelity in Fig.\ref{Fig2}(c)
changes drastically with model parameters.  It is because the band
edge states in the insulating phase II may be localized in
different regions of space and the overlap of these states may be
large or small.

\begin{figure}
\includegraphics[width=2.5in]{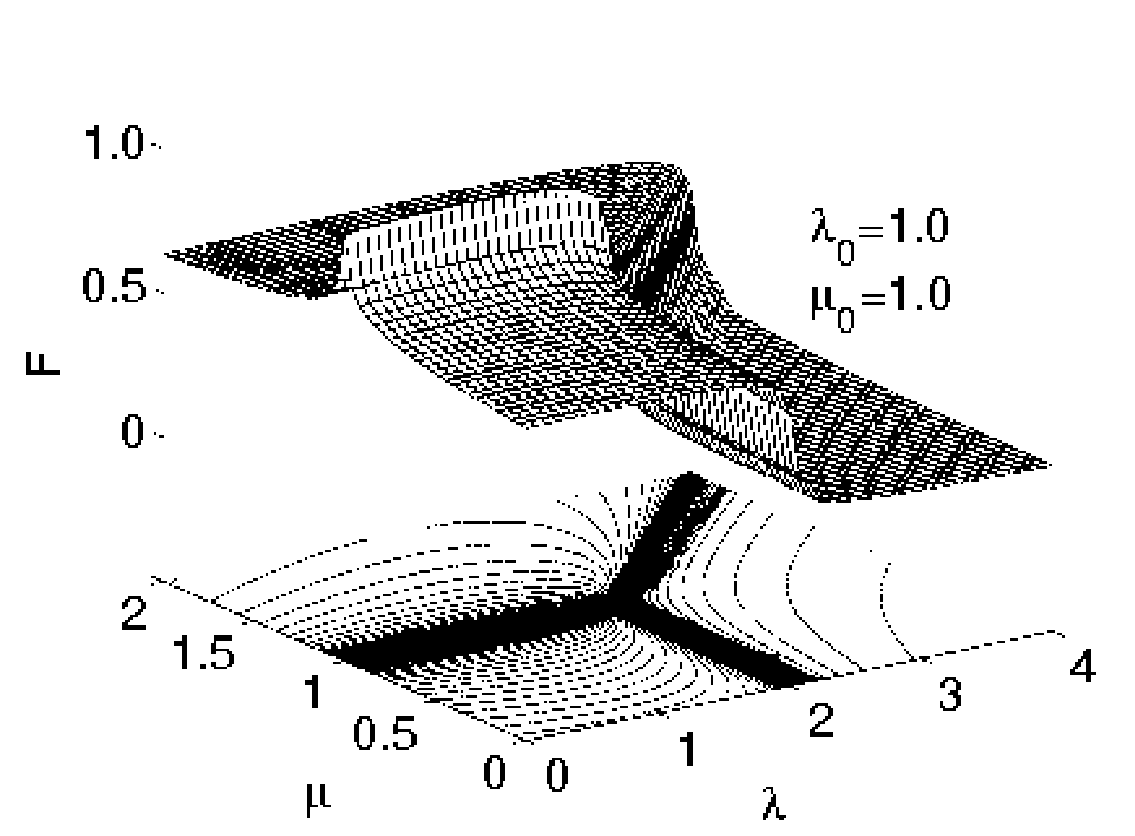}(a)
\includegraphics[width=2.5in]{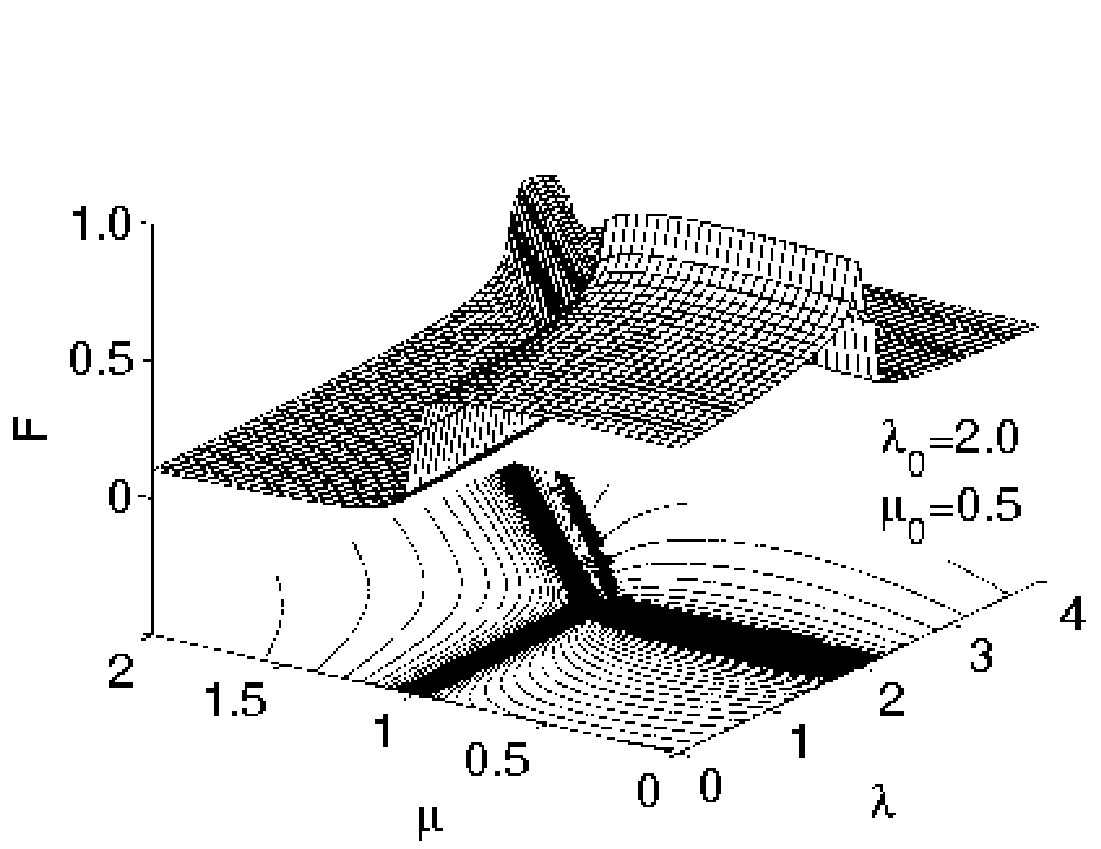}(b)
\includegraphics[width=2.5in]{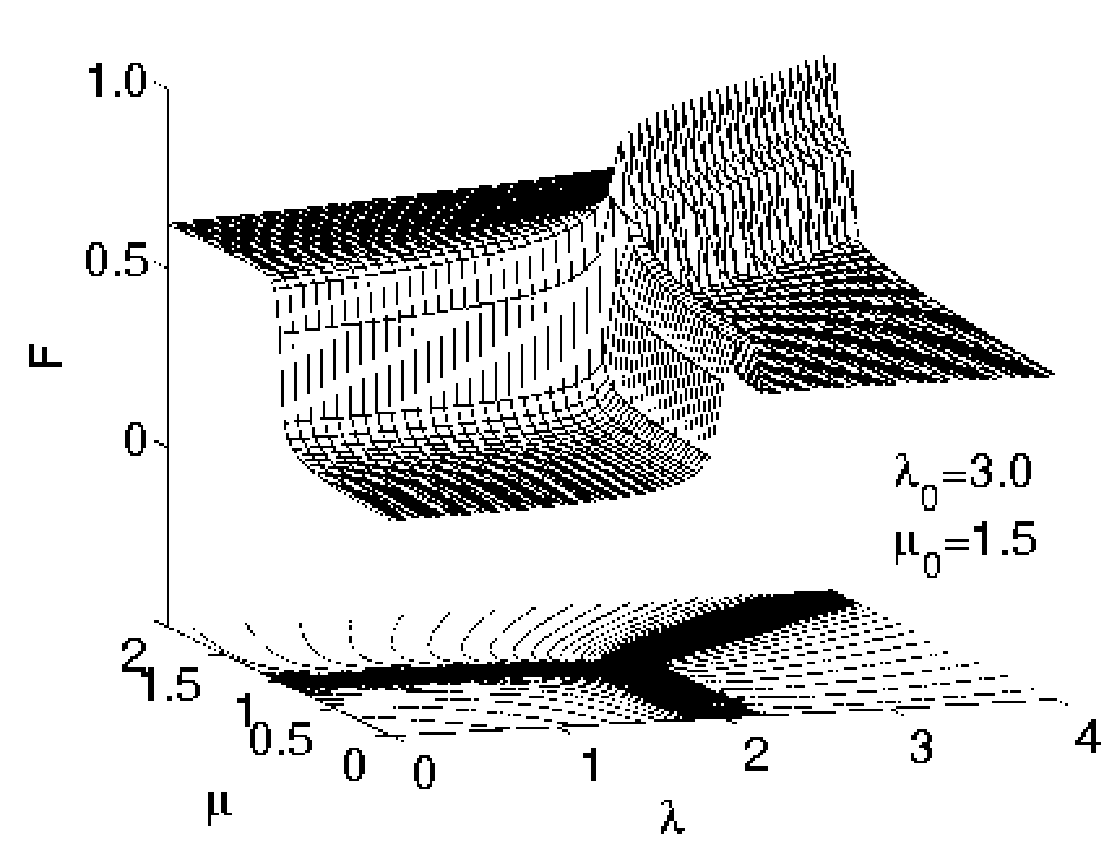}(c)
\caption{The fidelity $F(\lambda,\mu;\lambda_{0},\mu_{0})$ and its
contour map as functions of ($\lambda$,$\mu$) at (a), (b) and (c)
for ($\lambda_0,\mu_0$)=($1.0,1.0$), ($2.0,0.5$) and ($3.0,1.5$),
which corresponding to the system at the phase boundaries between
I and III,  I and II,  and III and II, respectively.}\label{Fig3}
\end{figure}

At the three phase boundaries, we choose
($\lambda_0,\mu_0$)=($1.0,1.0$), ($2.0,0.5$) and ($3.0,1.5$) as
examples, which corresponding to the system at the boundaries
between Phase I and III, Phase I and II and Phase III and II,
respectively. The fidelity $F$ varying with $\lambda$ and $\mu$
are shown in Fig.\ref{Fig3}. It shows that when parameters
($\lambda,\mu$) and ($\lambda_0,\mu_0$) are at a same critical
line, the fidelity is near one; otherwise, the fidelity is
relatively small.  It is interesting that if a point
($\lambda_0,\mu_0$) in the critical line between Phase I and
II(Phase I and II, Phase II and III), the fidelity in the both
phases is relatively large. Similar to that shown in
Fig.\ref{Fig2}, these contour maps of fidelity also divide the
parameter space to three regions, which is same as the phase
diagram shown in Fig.\ref{Fig1}.

\begin{figure}
\includegraphics[width=2.5in]{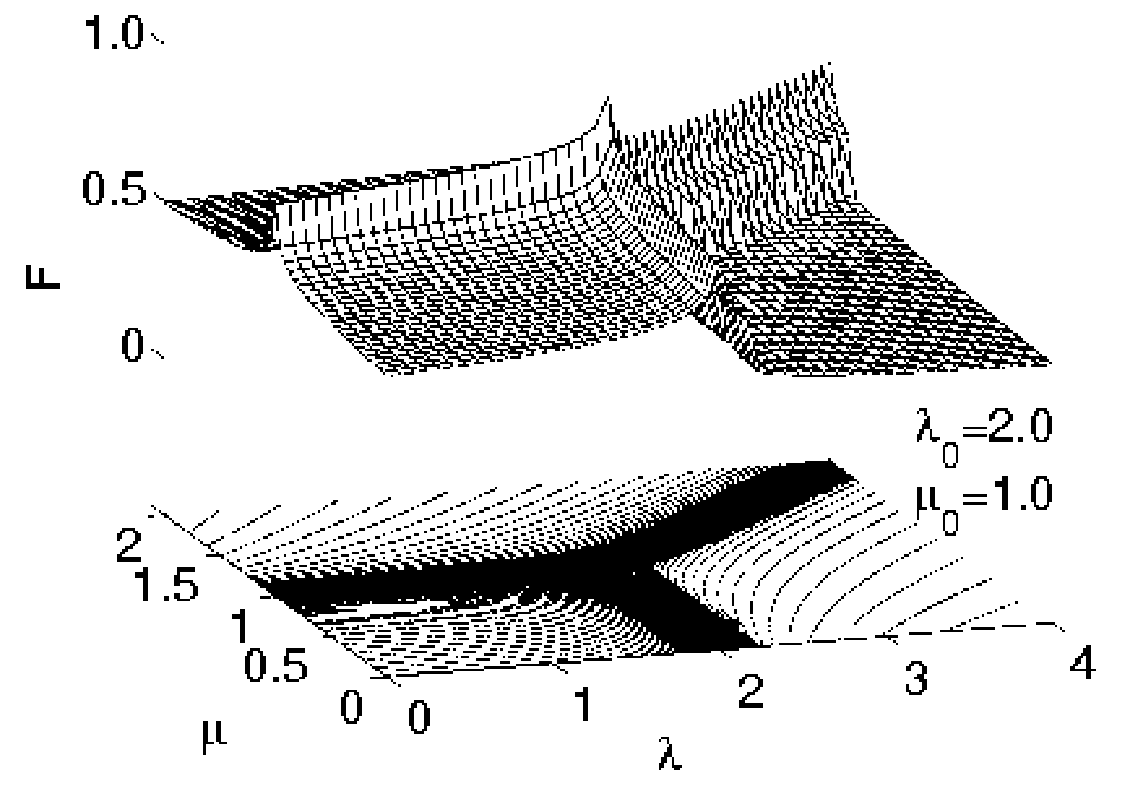}
\caption{The fidelity $F(\lambda,\mu;\lambda_{0},\mu_{0})$ and its
contour map as functions of ($\lambda$,$\mu$) for
($\lambda_0,\mu_0$)=($2.0,1.0$), which corresponding to the system
at the bicritical point.}\label{Fig4}
\end{figure}

At the bicritical point ($\lambda_0,\mu_0$)=($2.0,1.0$),  the
fidelity $F$ and its contour map as functions of $\lambda$ and
$\mu$ are plotted in Fig.\ref{Fig4}. It shows that when
($\lambda,\mu$)=($2.0,1.0$), $F$ is maximal and equal to one, when
($\lambda,\mu$) for the three critical lines, $F$ becomes
relatively small, and when ($\lambda,\mu$) in Phase I, II and III,
$F$ becomes relatively smaller. All these certify that the
bicritical point itself is different from others points, which is
agreement with the conclusion that the bicritical point is a
particular critical point as investigated in Ref.\cite{in06}.  At
the same time, the contour of fidelity can reflect the phase
diagram shown in Fig.\ref{Fig1}.

\subsection{\label{sec32} Fidelity Susceptibility}

\begin{figure}
\includegraphics[width=2.5in]{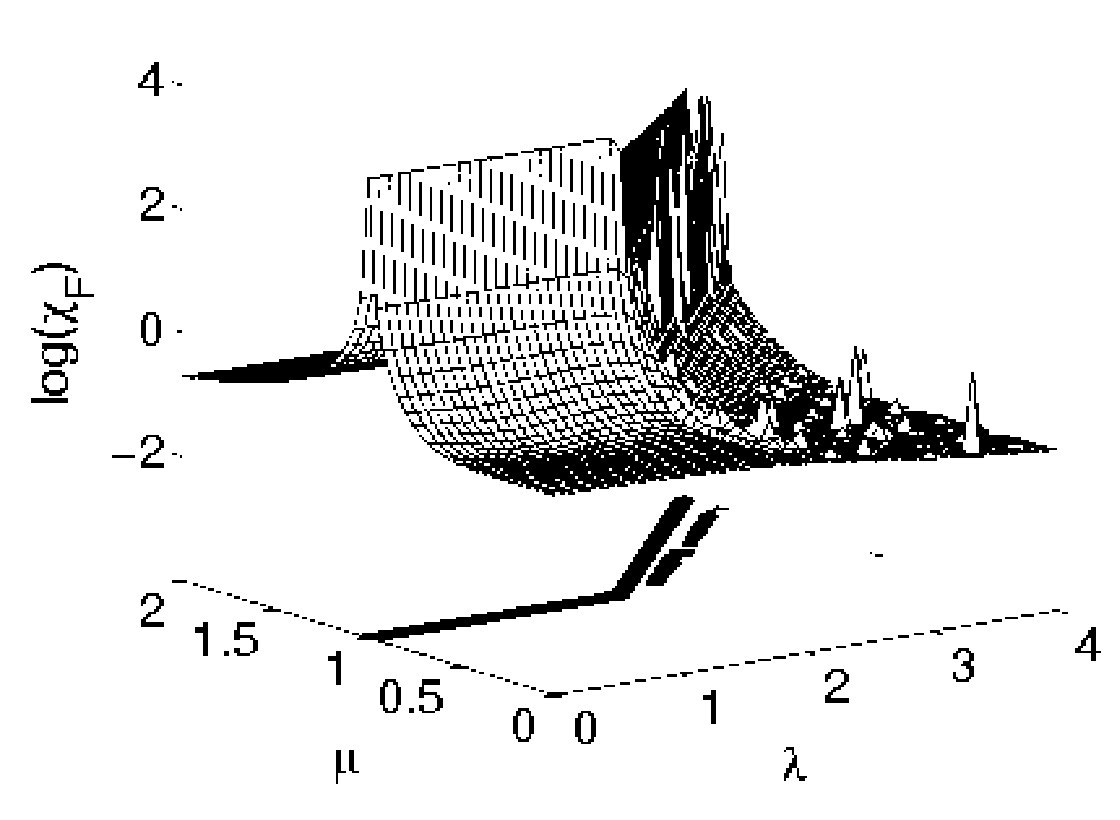}(a)
\includegraphics[width=2.5in]{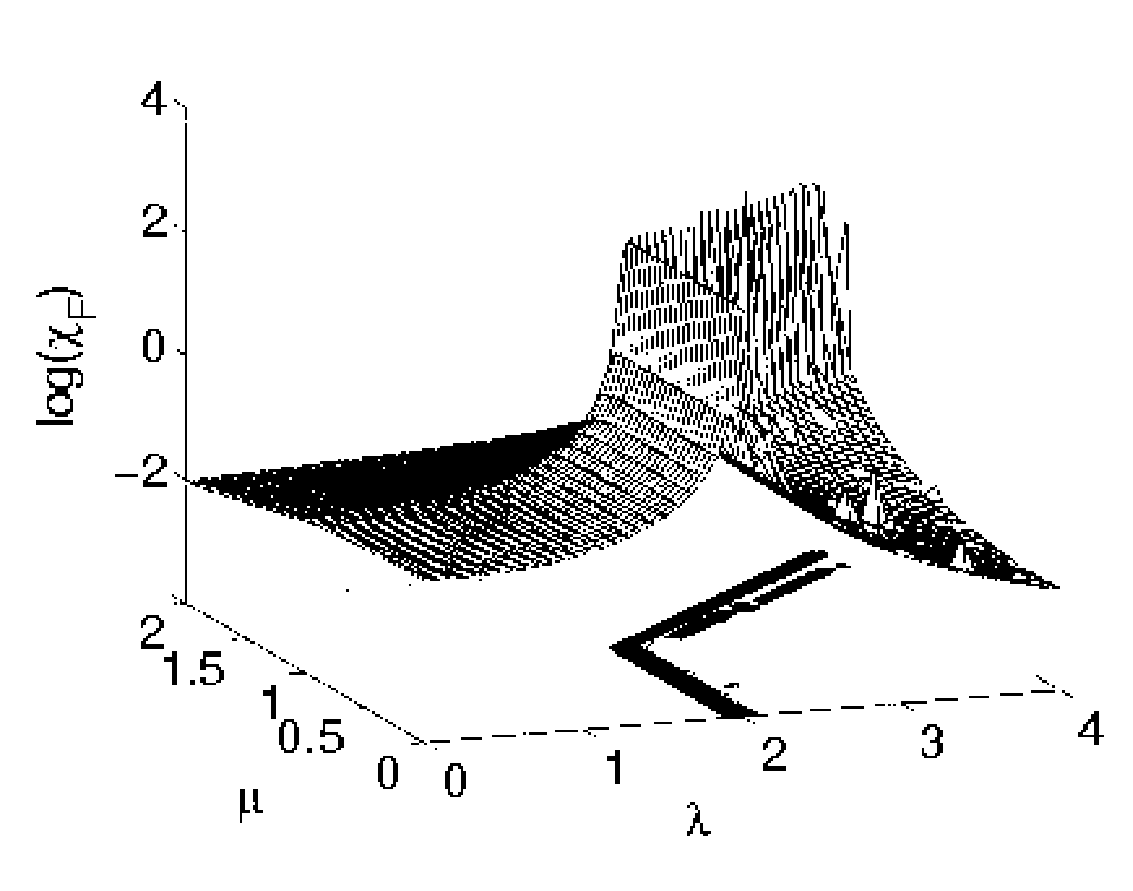}(b)
\includegraphics[width=2.5in]{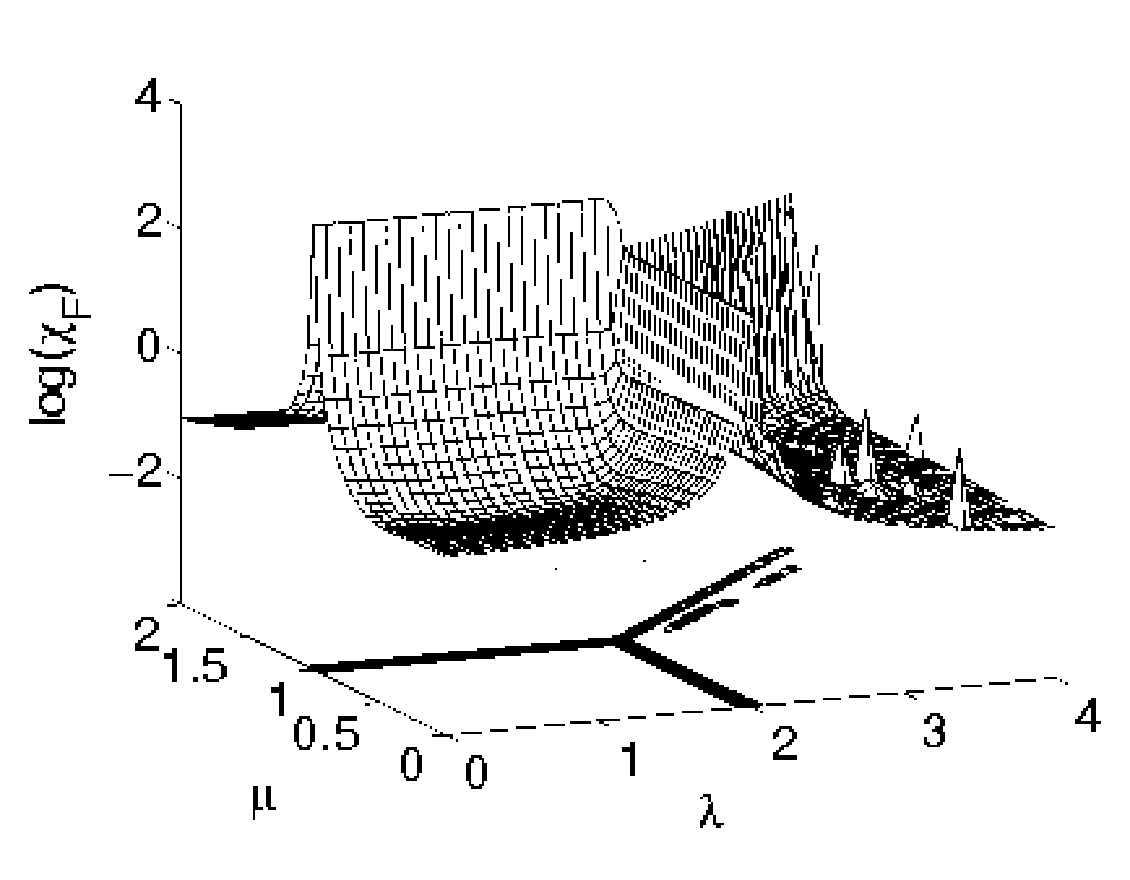}(c)
\caption{The logarithmic plots of fidelity susceptibility $\chi_F$
and its contour maps as functions of ($\lambda,\mu$) from
different parameter paths $q=q(\lambda,\mu)$. The tangent units
vector ($n^\lambda,n^\mu$) of paths equals to (a)($0,1$),(
b)($1,0$), (c)($1/\sqrt{5},-2/\sqrt{5}$), respectively.
}\label{Fig5}
\end{figure}

According to the definition of FS in Eq.(\ref{form6}), its values
depend on $g_{ab}$ and a specific direction of parameter path
$q=q(\lambda,\mu)$\cite{ya08}. The tangent unit vector
($n^\lambda,n^\mu$) which defines the direction may be different,
though $g_{ab}$ does not depend on parameter paths. In the
following, the FS for $(n^\lambda,n^\mu)=(0,1), (1,0)$, and
($1/\sqrt{5},-2/\sqrt{5}$) are shown in Fig.\ref{Fig5}(a), (b) and
(c), respectively, which are corresponding to that only $\mu$ ,
only $\lambda$ changes and both change simultanely.

\begin{figure}
\includegraphics[width=2.5in]{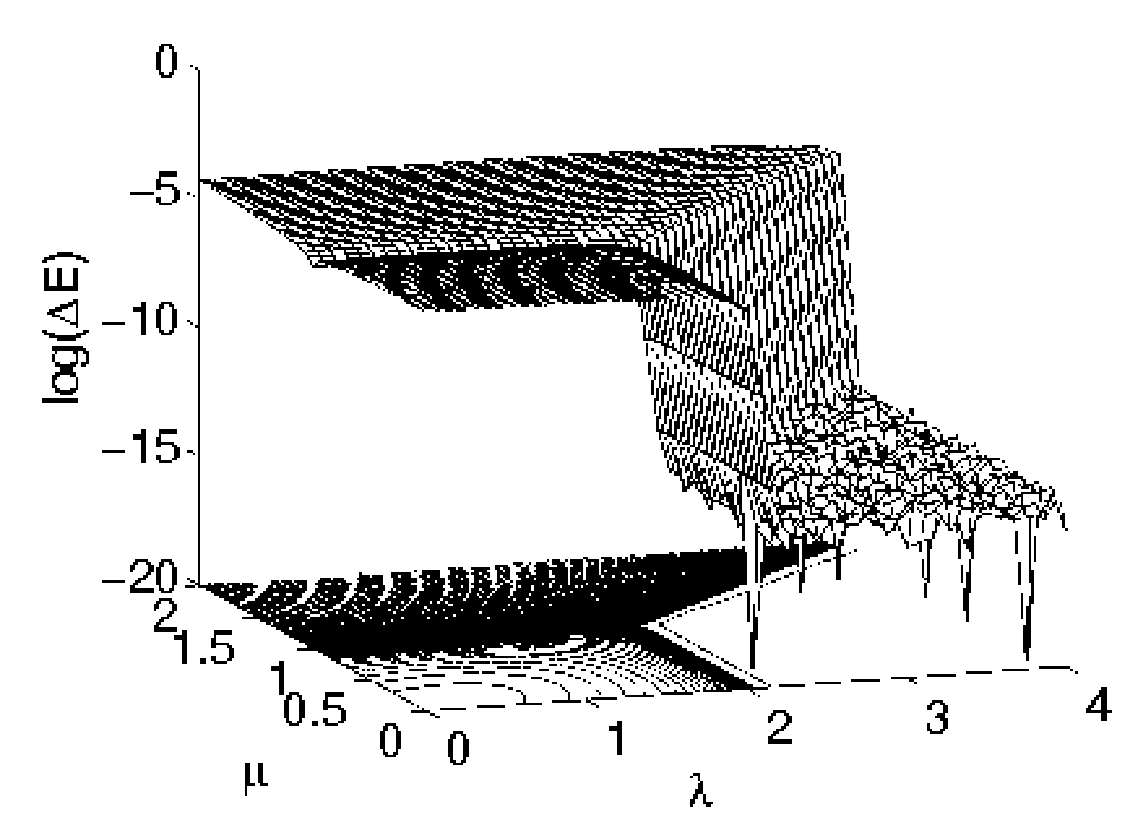}
\caption{ log ($\Delta E$) varying with ($\lambda$,$\mu$), here
$\Delta E$ is the gap between $E_1$ and $E_0$ at
($\lambda$,$\mu$).}\label{Fig6}
\end{figure}

Fig.\ref{Fig5} shows the $\chi_F(\lambda,\mu)$ are different when
choosing different parameter paths. For Fig.\ref{Fig5}(a), only the
driving Hamiltonian $H_\mu$ effects the values of $\chi_F$. From the
corresponding contour maps, the boundaries between the metallic
phase III and other two phases are identified, i.e., there are sharp
changes in $\chi_F$ at these phase boundaries, while for
Fig.\ref{Fig5}(b), only $H_\lambda$ effects $\chi_F$ and the
boundaries between the insulating phase II and other two phases are
identified. The combination of the two contour maps in
Fig.\ref{Fig5}(a) and (b) is consistence with  the phase diagram
that shown in Fig.\ref{Fig1}. For Fig.\ref{Fig5}(c), both
$H_\lambda$ and $H_\mu$ effect $\chi_F$ and the corresponding
contour maps itself can reflect the phase diagram.  In
Fig.\ref{Fig5}(a), (b) and (c), the varying of $\chi_F$ in the
insulating phase II is not smooth, which is due to the gap between
$E_0$ and $E_\gamma$ may be close to zero at some parameters(see
Eq.\ref{form9}). For this, the logarithmic plots of the gap $\Delta
E$ for the first excited state eigenenergy $E_1$ and ground state
eigenenergy $E_0$ varying with ($\lambda,\mu$) are shown in
Fig.\ref{Fig6}. One sees that in Phase II, all the values of $\Delta
E$ are very small and some almost are equal to zeros, therefore the
fidelity $F$ changes sharply at these parameter points. It is
interesting that the contour maps of $\Delta E$ divide the parameter
space to three regions, which is also consistence with the phase
diagram shown in Fig.\ref{Fig1}.

\begin{figure}
\includegraphics[width=2.5in]{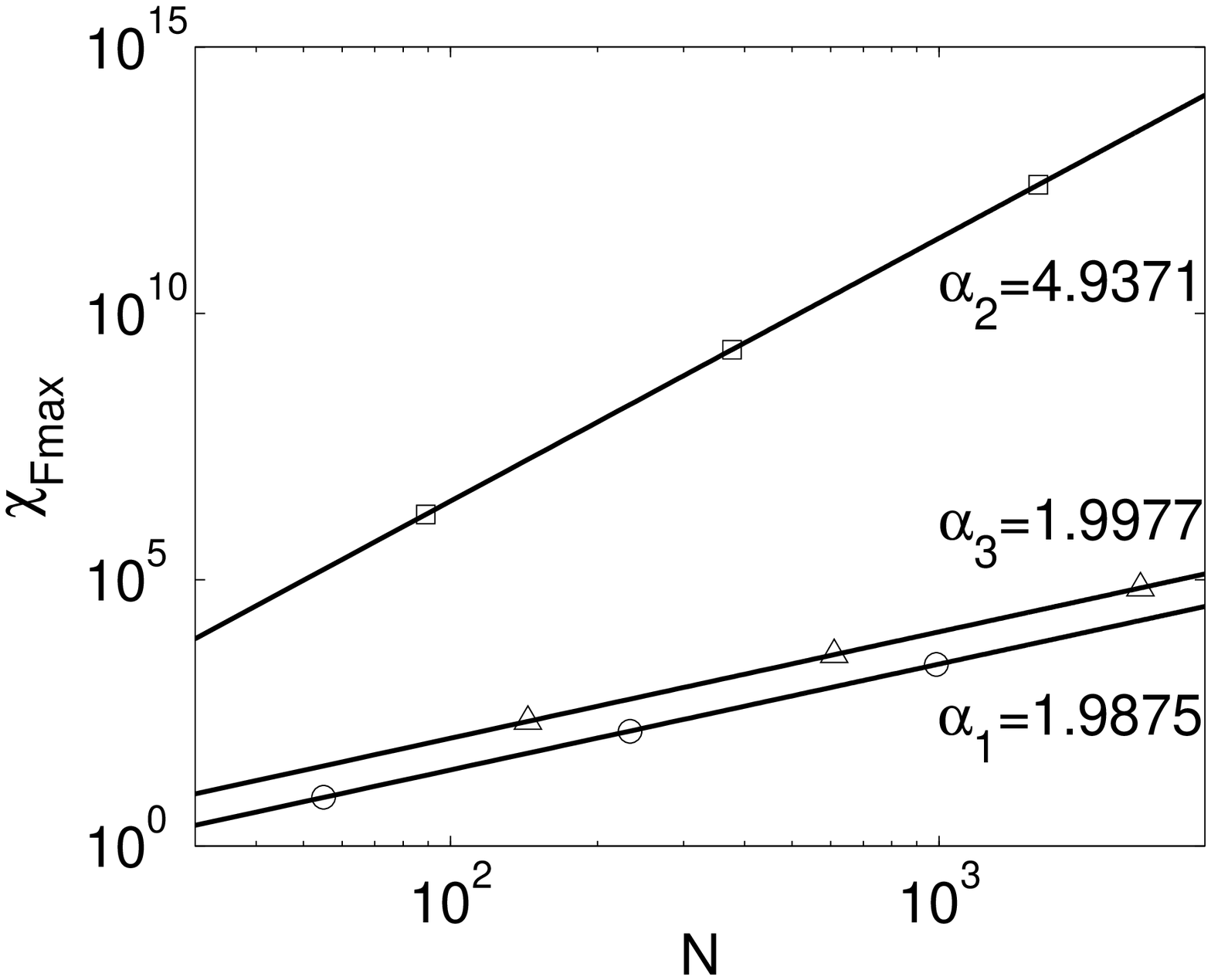}(a)
\includegraphics[width=2.5in]{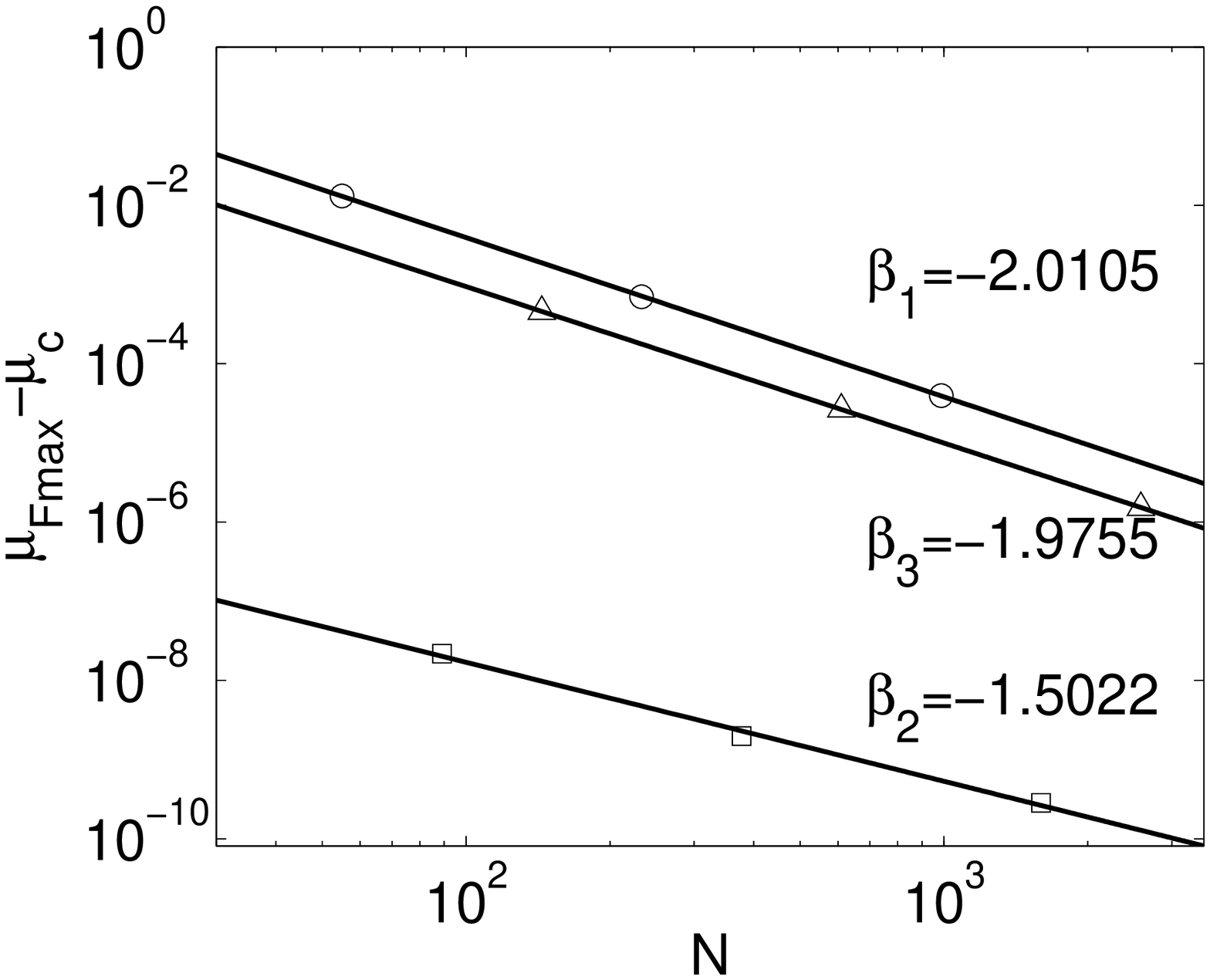}(b)
\caption{The scaling behaviors of (a) $\chi_{Fmax}$ and (b)
$\mu_{Fmax}-\mu_c$, respectively. The system sizes $F_{3l}=55,
233, 987$($\circ$),$F_{3l+1}=89, 377, 1597$($\square$) and
$F_{3l+2}=144, 610, 2584$($\triangle$), respectively. At the same
time, the corresponding fitted lines are also shown, respectively.
}\label{Fig7}
\end{figure}

\begin{figure}
\includegraphics[width=2.5in]{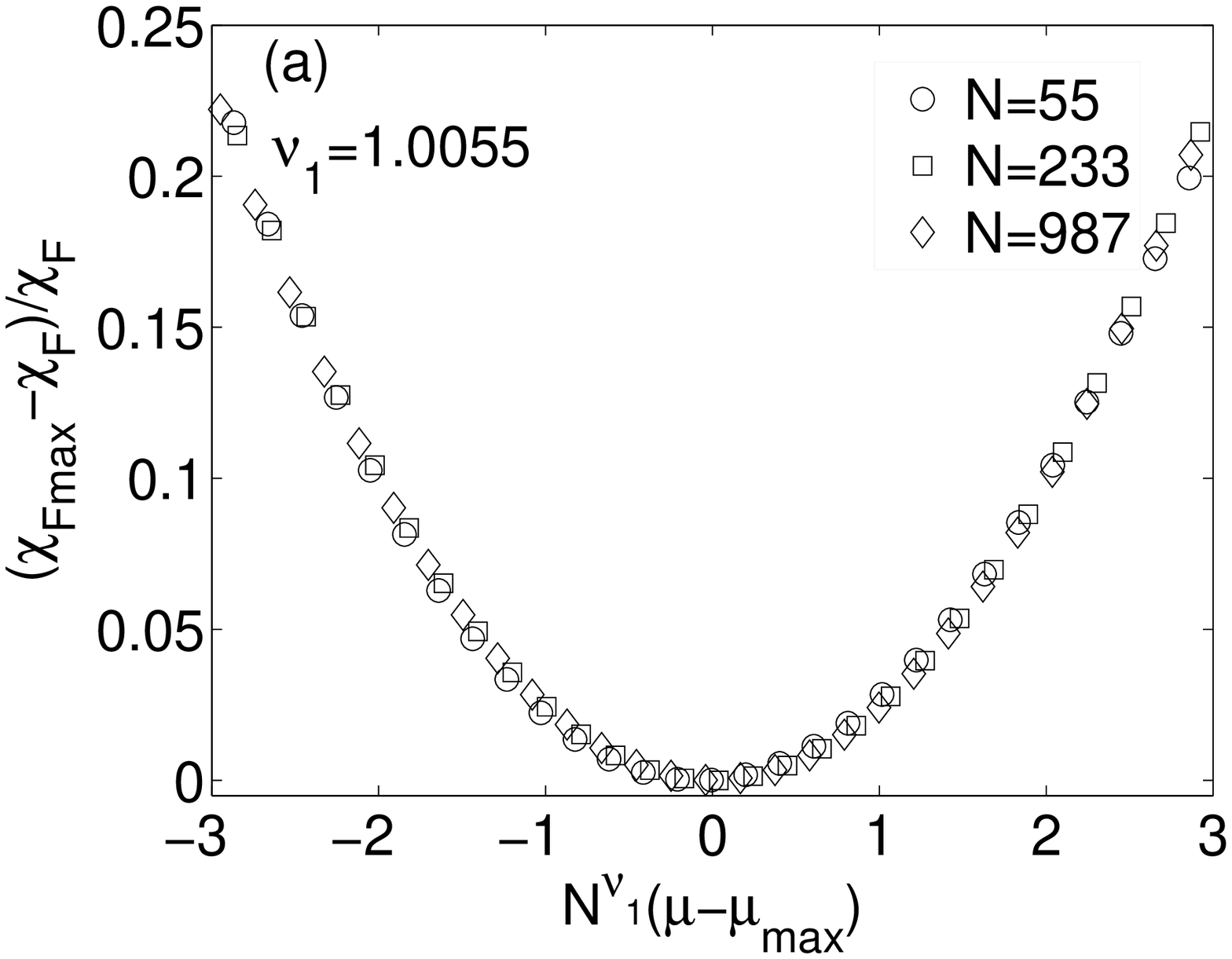}(a)
\includegraphics[width=2.5in]{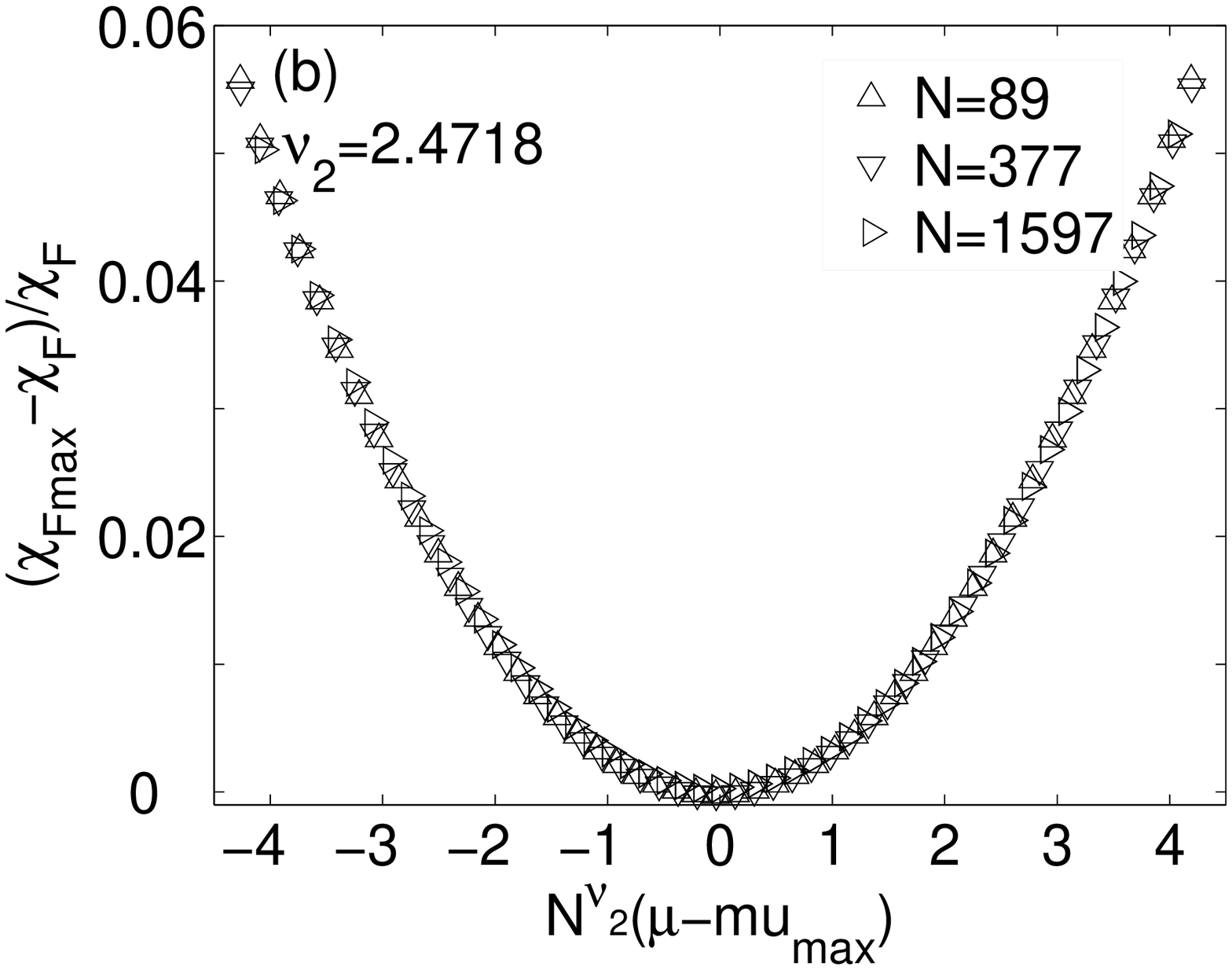}(b)
\includegraphics[width=2.5in]{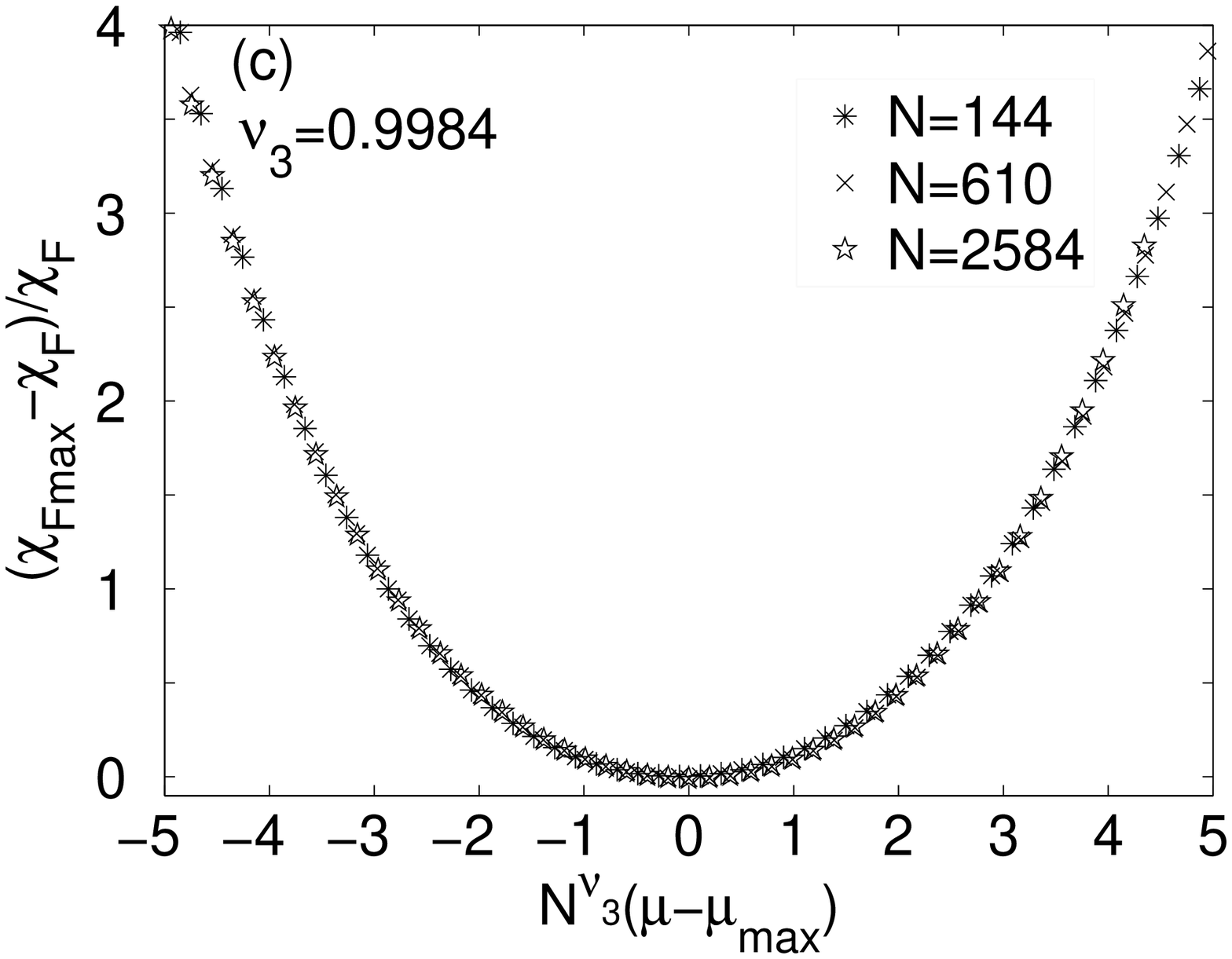}(c)
\caption{The finite size scaling analysis is performed for
$(\chi_{Fmax}-\chi_F)/\chi_F$ as a function
$N^\nu(\mu-\mu_{max})$. }\label{Fig8}
\end{figure}

In order to study the critical behavior around critical
points($\lambda_c,\mu_c$), we study the finite scaling analysis of
FS\cite{yo07,gu08,kw07,ya08} and obtain the corresponding critical
exponents. It has been found that these critical exponents and
different scaling behaviors of FS can characterize the
universality classes of phase transitions\cite{gu08}. Firstly, we
study the transition between the metallic phase I and metallic
phase III and choose the critical point
($\lambda_c=1.0,\mu_c=1.0$) as an example. Near the critical
parameter with the tangent units vector ($n^\lambda=0,n^\mu=1$) of
parameter paths, the $\chi_F$ is calculated for various system
sizes $N$, which corresponding to the case that shown in
Fig.\ref{Fig6}(a). Along the parameter path, the FS reaches its
maximum value $\chi_{Fmax}$ at a certain position $\mu_{max}$. The
scaling behaviors of $\chi_{Fmax}$ and $\mu_{Fmax}-\mu_c$ are
given in Fig.\ref{Fig7}(a) and (b), respectively, which shows that
$\chi_{Fmax}\varpropto N^\alpha$ and $\mu_{Fmax}-\mu_c\varpropto
N^\beta$. All $\beta<0$, which means $\mu_{Fmax}$ tends to the
critical point $\mu_c$ in the thermodynamic limit. For the system
sizes $N$ are chosen to the Fibonacci number $F_m$ with $m=3l+1$
and $m\neq3l+1$ for integer $l$, the system sizes are divided to
two cases\cite{in06}. It is found that $\alpha\approx2.0$ and
$\beta\approx-2.0$ for $m\neq3l+1$, while $\alpha=4.9371$ and
$\beta=-1.5022$ for $m=3l+1$. In Fig.\ref{Fig8}, the corresponding
scaling functions are plotted. It shows that the exponent
$\nu\approx1.0$ for $m\neq3l+1$, while $\nu=2.4718$ for $m=3l+1$.
Although $\alpha$($\beta$)are different for $m=3l+1$ and
$m\neq3l+1$, the scaling relation $\alpha/\nu\approx2$ is
universal. The scaling relation is same as that for a
one-dimensional asymmetric Hubbard model studied by Gu \textit{et
al}\cite{gu08}.

To understand the different behaviors between the systems with
$N=F_{3l+1}$ and $N\neq F_{3l+1}$, we analyse carefully the
structure of the system. According to Fibonacci numbers
$F_m=F_{m-2}+F_{m-1}$ with $F_0=F_1=1$, $F_{3l}$ and  $F_{3l+1}$ are
odd, which can be written as ${2k_1+1}$ and ${2k_2+1}$ with integers
$k_1$ and $k_2$, respectively. For $N=F_{3l+1}$,
$\phi=\frac{F_{3l}}{F_{3l+1}}=\frac{2k_1+1}{2k_2+1}$ and $\mu=1$,
the hopping term of Eq.(\ref{form3}), $-[1+\mu e^{ 2\pi
i\phi(n+1/2)}]=-[1+e^{2\pi i\frac{2k_1+1}{2k_2+1}\frac{2n+1}{2}}]=0$
at the site $n=k_2$, i.e., a bond between \textit{$k_2$} and
(\textit{$k_2+1$})th site breaks. The system is divided to two
segments. For $N\neq F_{3l+1}$, it does not happen. This induces
differences between the energy spectrum of $N=F_{3l+1}$ and $N\neq
F_{3l+1}$ \cite{in06}. 

\begin{figure}
\includegraphics[width=2.5in]{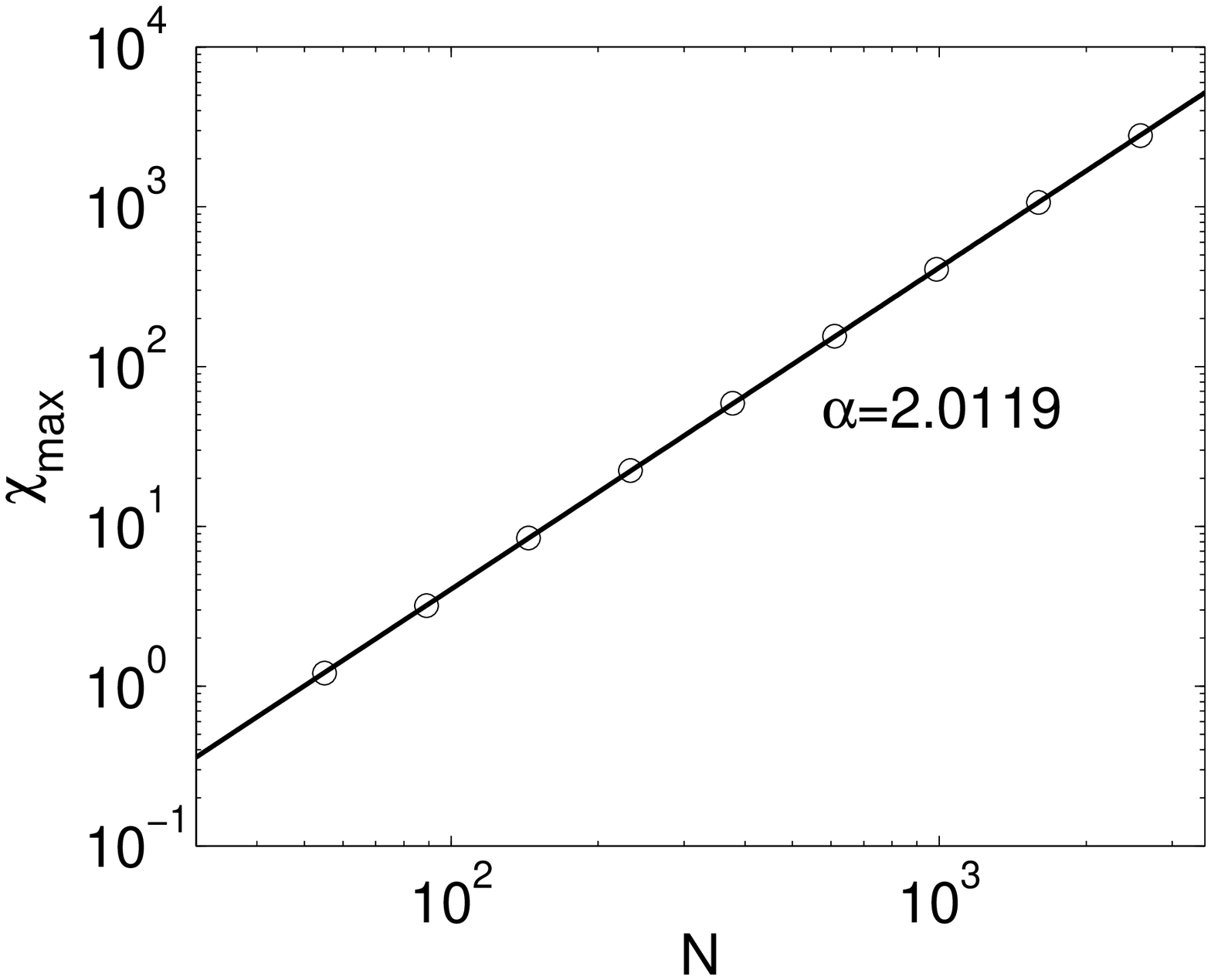}(a)
\includegraphics[width=2.5in]{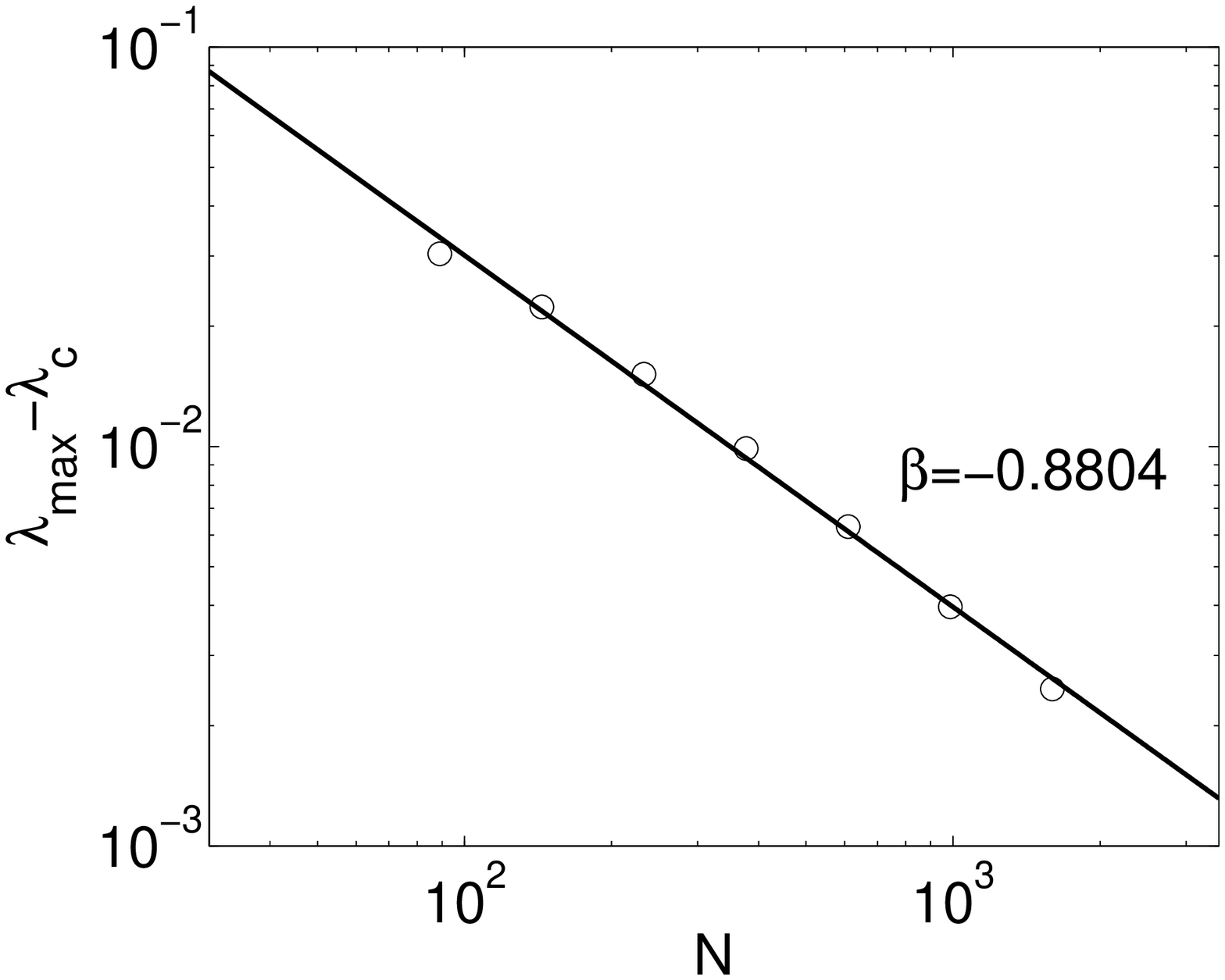}(b)
\caption{The scaling behaviors of (a) $\chi_{Fmax}$ and (b)
$\lambda_{Fmax}-\lambda_c$, respectively. The system sizes are
$55, 89, ...,2584$.}\label{Fig9}
\end{figure}

\begin{figure}
\includegraphics[width=2.5in]{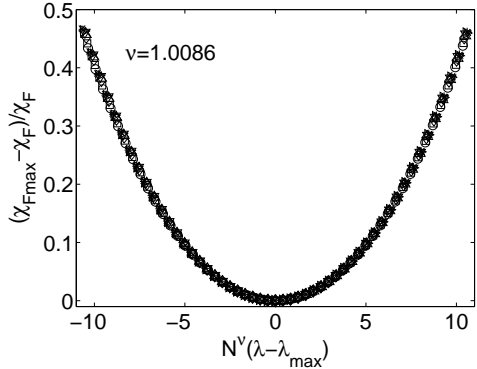}
\caption{The finite size scaling analysis is performed for
$(\chi_{Fmax}-\chi_F)/\chi_F$ as a function
$N^\nu(\lambda-\lambda_{max})$ for the system sizes are $55, 89,
...,2584$.}\label{Fig10}
\end{figure}

Secondly, we study the transition between the metallic phase I and
insulating phase II and choose the critical parameter
($\lambda_c=2.0,\mu_c=0.5$) as an example. Near the critical
parameter with the tangent units vector ($n^\lambda=1,n^\mu=0$),
the FS $\chi_F$ is calculated for various system sizes $N$, which
corresponding the case shown in Fig.\ref{Fig6}(b). From Figs.
\ref{Fig9} and \ref{Fig10}, it is found that for all system sizes,
$55, 89, ...,2584$, the exponents $\alpha,\beta$ and $\nu$ are
same and the the scaling relation $\alpha/\nu\approx2$ is also
obtained. We have studied the transition between the metallic
phase III and insulating phase II, the results are similar and the
relation $\alpha/\nu\approx2$ is also tenable.

\subsection{\label{sec33} von Neumann entropy}

\begin{figure}
\includegraphics[width=2.5in]{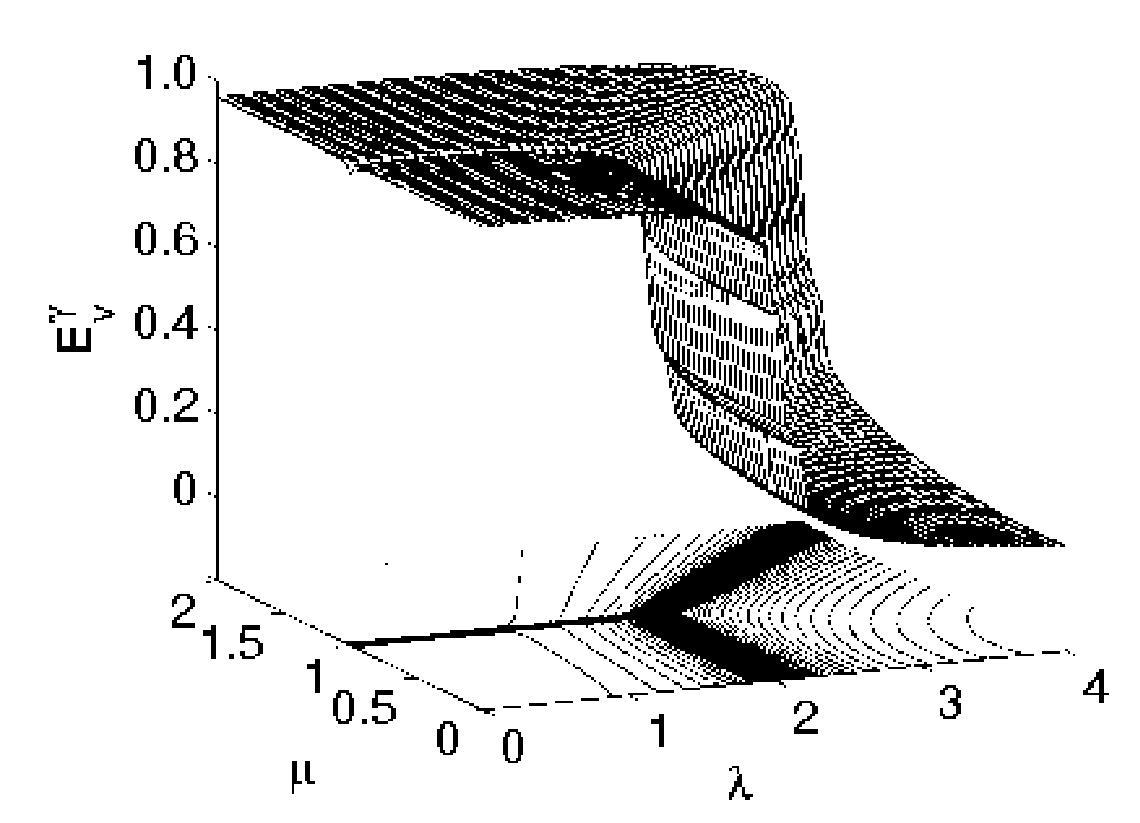}(a)
\includegraphics[width=2.5in]{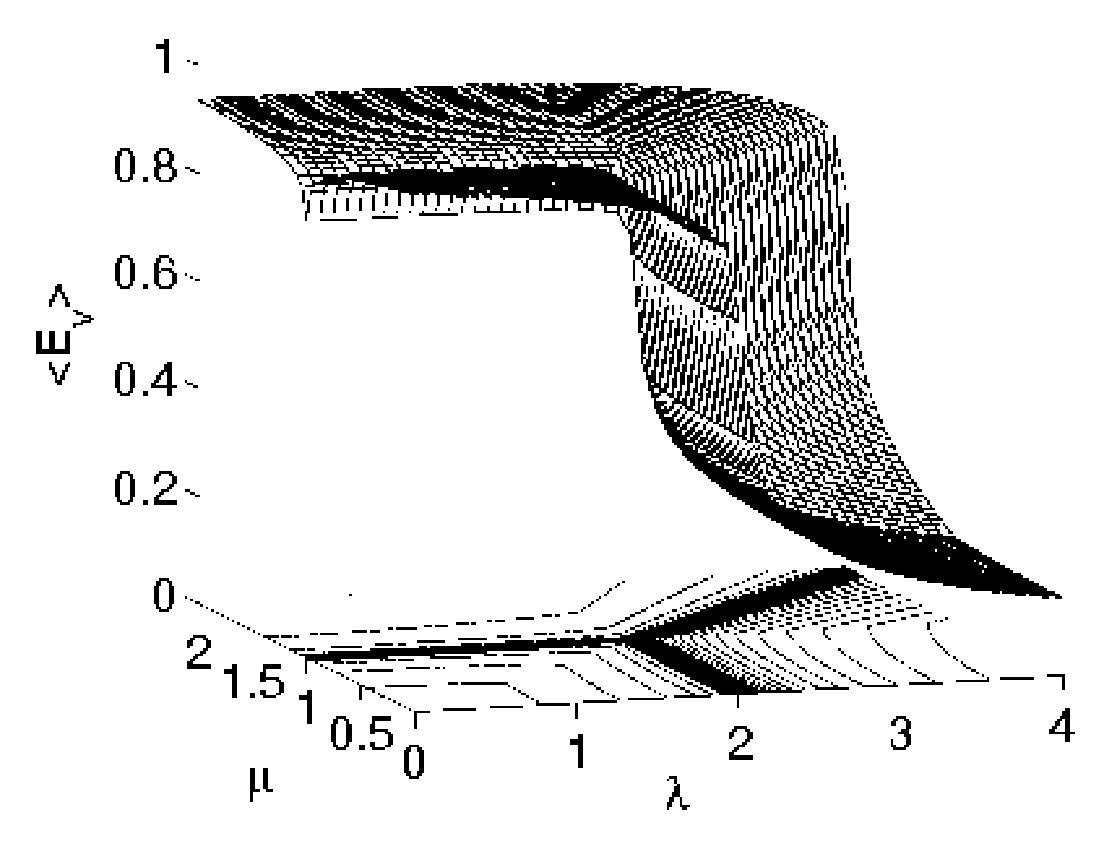}(b)
\caption{The site-averaged von Neumann entropy $E^\gamma_v$ for
the lowest edge states (a) and the spectrum-averaged von Neumann
entropy $\langle E_v\rangle$ (b)as as functions of
($\lambda,\mu$), respectively.}\label{Fig11}
\end{figure}
The von Neumann entropy has been found to be a suitable quantity
to characterize the localization properties of electronic
states\cite{go05,go06,go07}. Fig.\ref{Fig11}(a) and (b) show the
site-averaged von Neumann entropy $E^\gamma_v$ for the lowest edge
states and the spectrum-averaged von Neumann entropy $\langle
E_v\rangle$, respectively. The varyings of the two quantities with
parameters($\lambda,\mu$) are similar. $E^\gamma_v$($\langle
E_v\rangle$) is near $1$ in the metallic phase I and III and
relatively small in the insulating phase II. There are sharp
decreases in $E^\gamma_v$($\langle E_v\rangle$) at phase
boundaries. The contour maps of them  divide the parameter space
to three parts, which is consistence with the phase diagram shown
in Fig.\ref{Fig1}.

\begin{figure}
\includegraphics[width=2.5in]{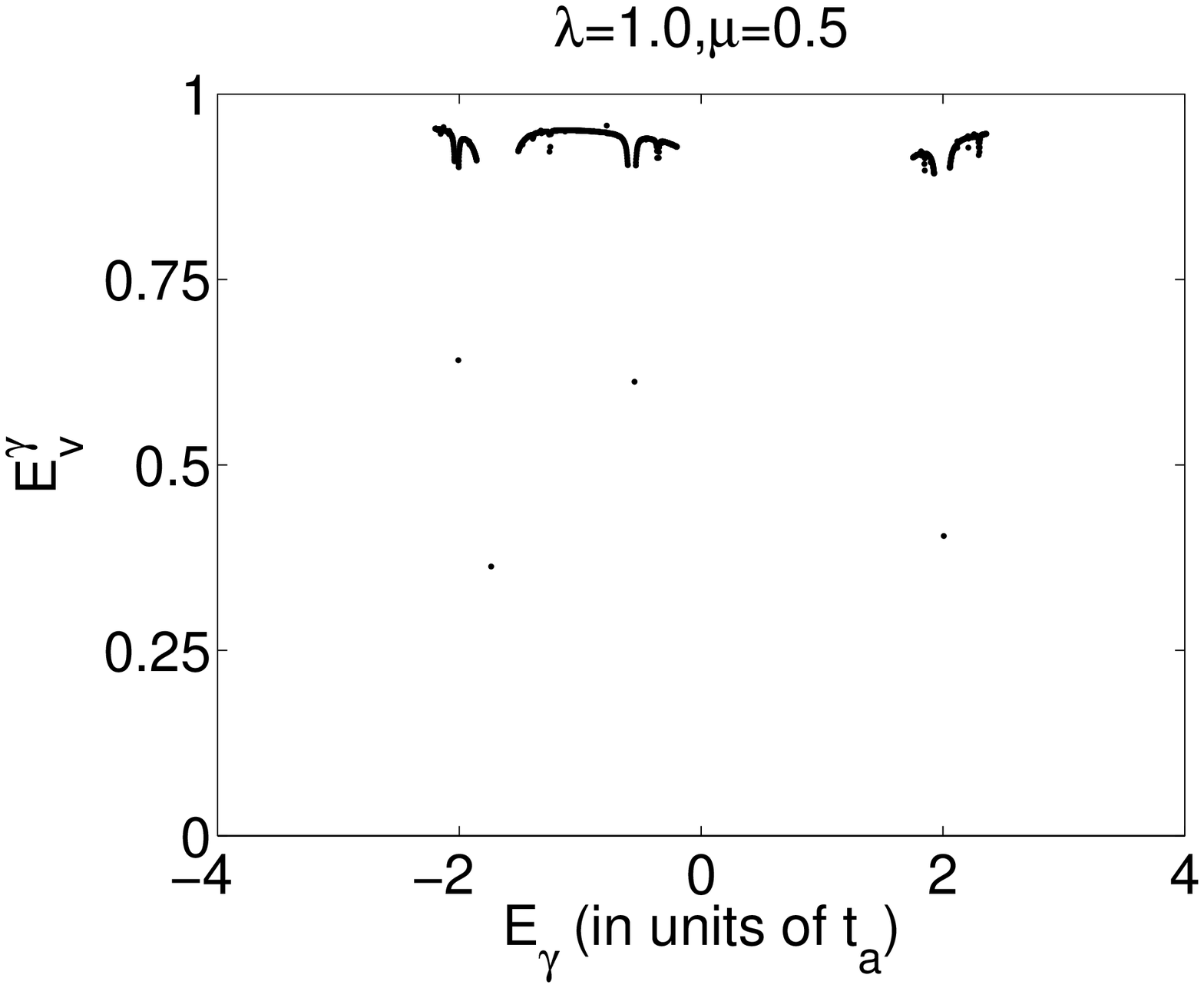}(a)
\includegraphics[width=2.5in]{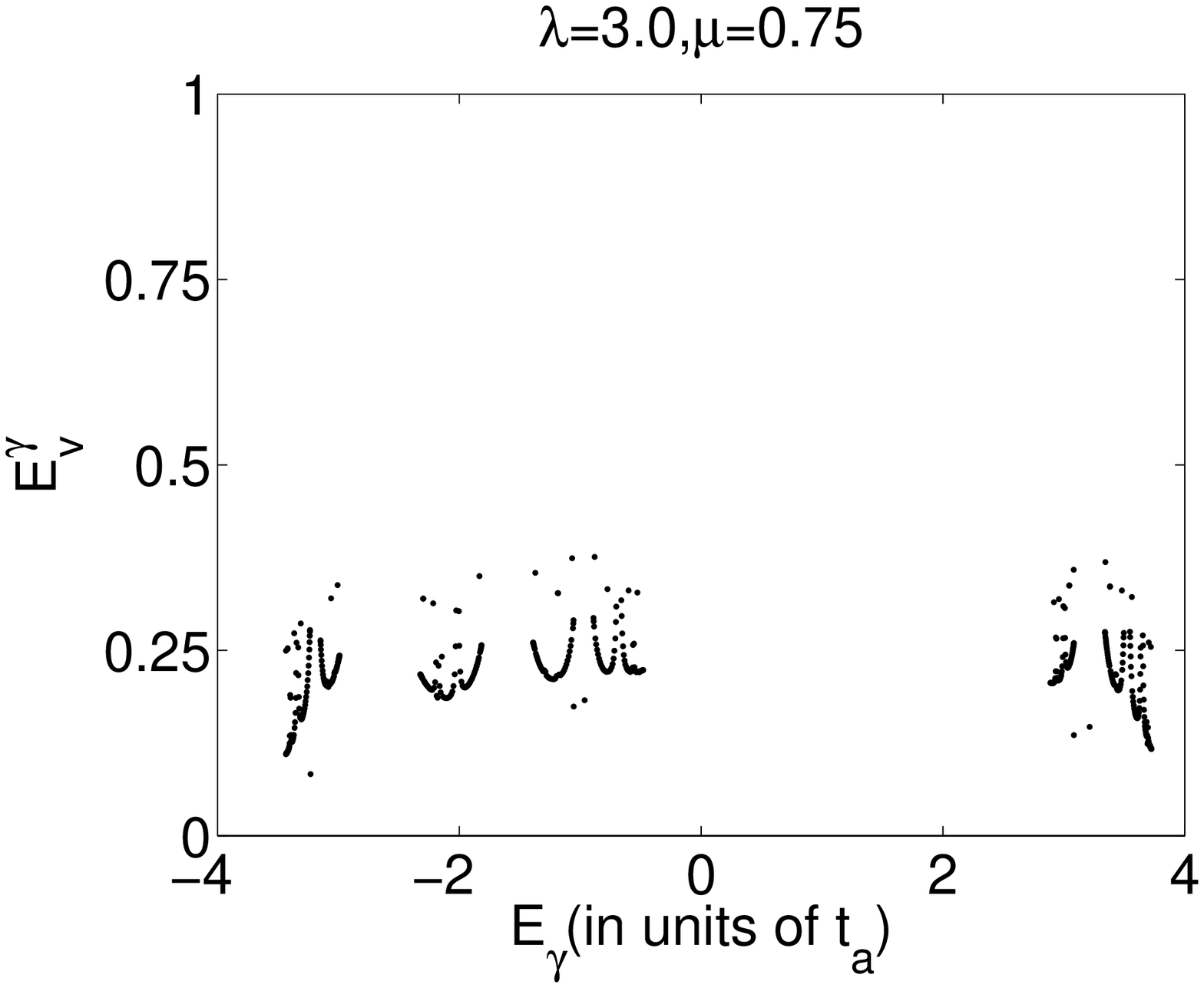}(b)
\includegraphics[width=2.5in]{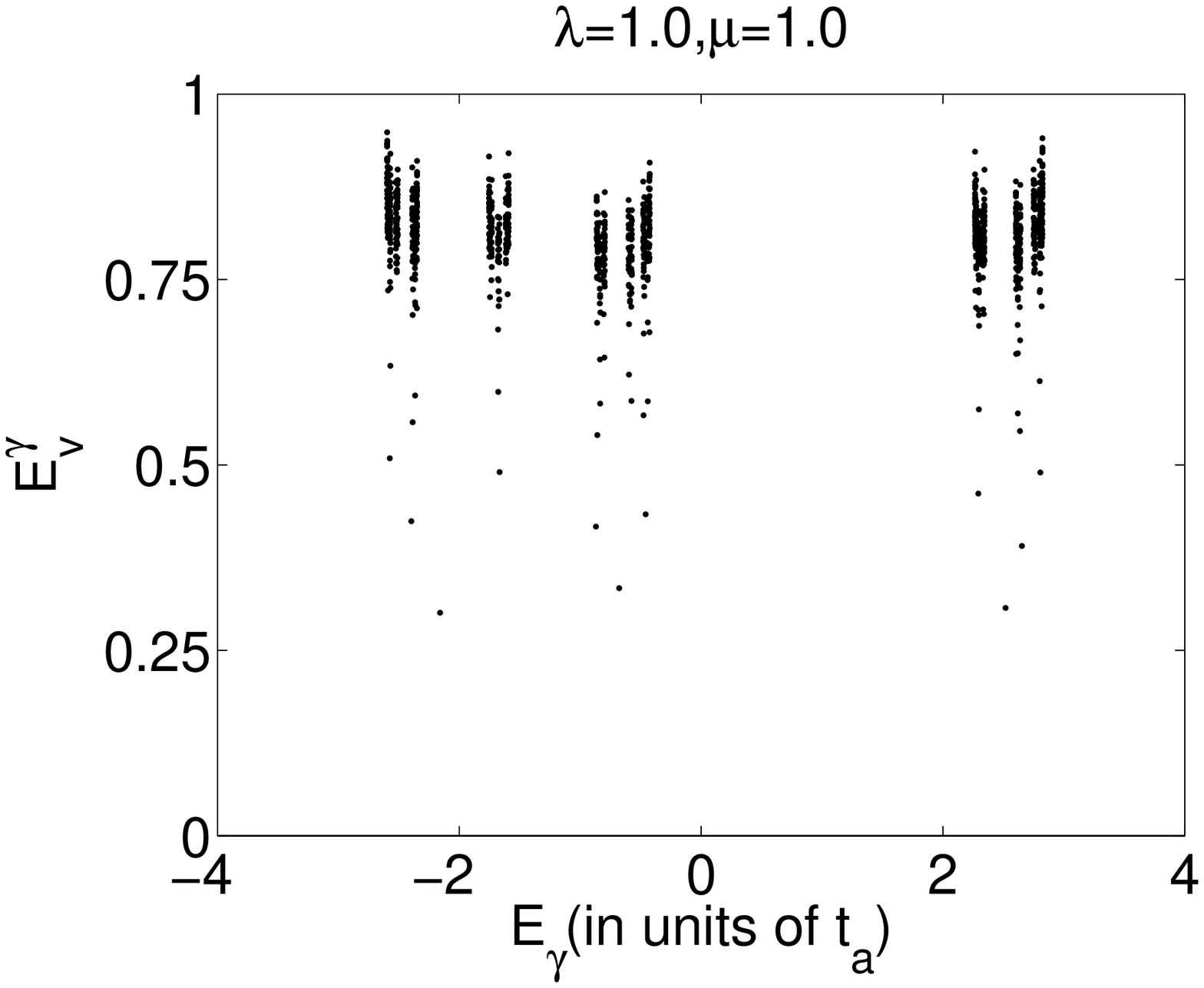}(c)
\includegraphics[width=2.5in]{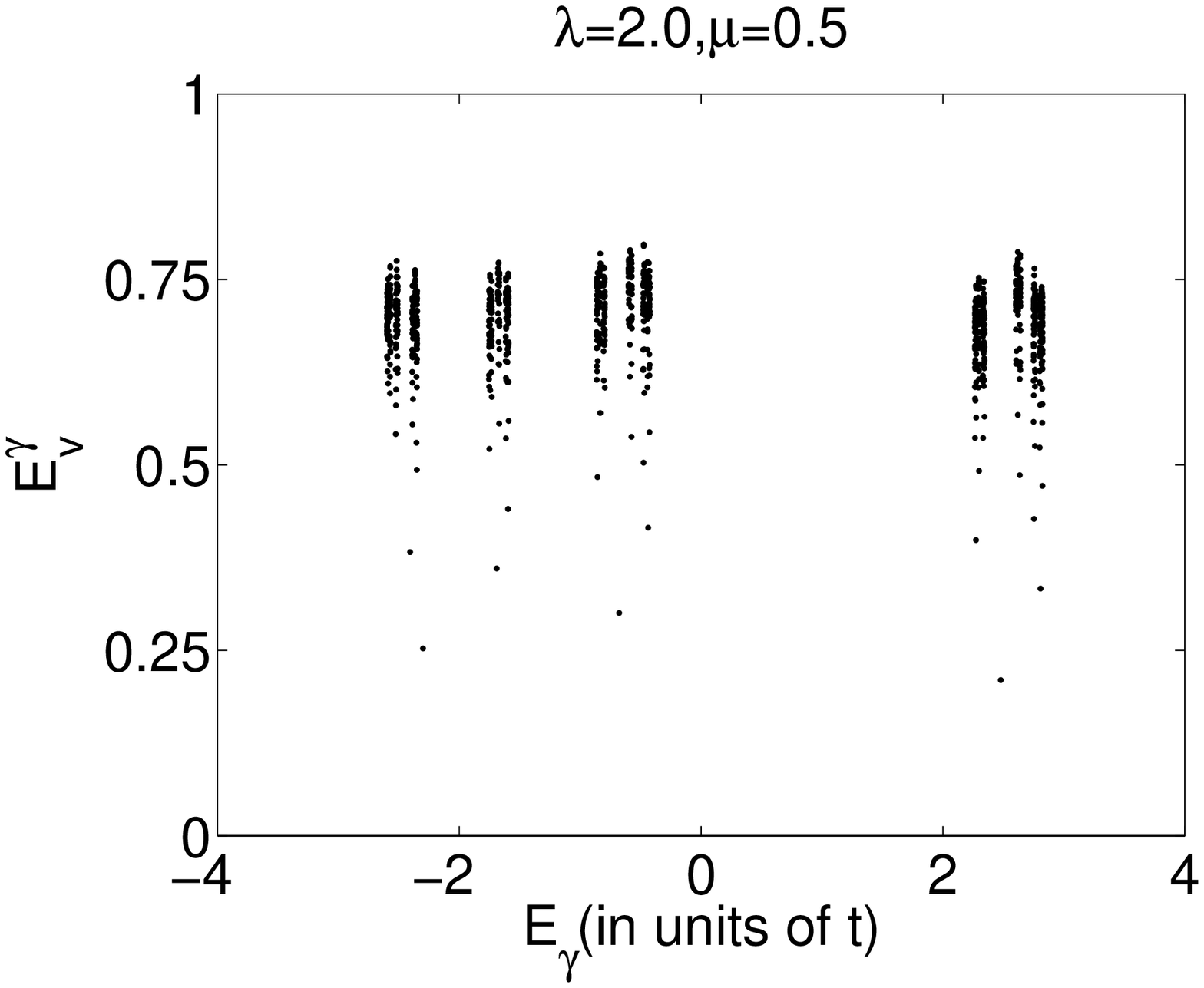}(d)
\caption{The site-averaged von Neumann entropy $E^\gamma_v$
 varying with eigenenergy $E_\gamma$ at
$(\lambda,\mu)=(1.0,0.5), (3.0,0.75), (1.0,1.0)$ and $(2.0,1.0)$
for (a), (b), (c) and (d), respectively.}\label{Fig12}
\end{figure}

Conventionally, the inverse participation ratio (IPR) is often
used as a measure of the wave-functions localization
length\cite{kr93}. The larger is the IPR, the more delocalized the
eigenstate is. It has found that the site-averaged von Neumann
entropy $E^\gamma_v$ increases exponentially with the IPR
\cite{go07}, i.e., $E^\gamma_v$ can reflect the localization
properties of electronic states. Fig.\ref{Fig12}(a)-(d) show
$E^\gamma_v$ varying with eigenenergy $E_\gamma$ for
$(\lambda,\mu)=(1.0,0.5), (3.0,0.75), (1.0,1.0)$ and $(2.0,0.5)$,
which corresponding to the metallic phase, the insulating phase,
the MMT and the MIT, respectively. In Fig.\ref{Fig12}(a), almost
all the $E^\gamma_v$ are near $1$, which means these states are
delocalized. Comparing Fig.\ref{Fig12}(b) with Fig.\ref{Fig12}(a),
all the $E^\gamma_v$ are small, which means that all eigenstates
are localized. In Fig.\ref{Fig12}(c) and (d), there coexist large,
middle, and small $E^\gamma_v$, which means the eigenstates are
critical. Though all eigenstates for the three phase boundaries
and the bicritical point are critical, the values of the
spectrum-averaged von Neumann entropy $\langle E_v\rangle$ are
different. Comparing to each other, the $\langle E_v\rangle$ for
boundaries between the metallic phase I and III are large, for the
bicritical point are middle, and for boundaries between metallic
phase and insulator are small, which can be seen from
Fig.\ref{Fig11}(b). All these indicate that, judging from the
varying of von Neumann entropy with parameter $(\lambda,\mu)$, the
phase diagram can be completely characterized.

\section{\label{sec4}Conclusions and Discussions}
For the extended Harper model introduced in Ref.\cite{in06}, we
have studied the fidelity between two lowest band edge states
corresponding to different model parameters, the FS and the von
Neumann entropy of the lowest band edge states, and the
spectrum-averaged von Neumann entropy. All the three quantities
can well characterize the rich phase diagram of the interesting
model.

In detail, firstly, the fidelity varying with parameters
($\lambda,\mu$) for seven groups of fixing values
($\lambda_{0},\mu_{0}$) is studied, which corresponding to
different phases, different phase boundaries and the bicritical
point. When parameters are in the same phase or same boundary, the
fidelity is near one, otherwise, it is small. There are drastic
changes in fidelity when one parameter corresponding to phase
boundaries. At the same time, the contour maps of fidelity divide
the parameter space to three regions, which is a good agreement
with the phase diagram of the model. In fact, these conclusions
are valid for arbitrary fixing values($\lambda_{0},\mu_{0}$) in
the parameter space. It indicates that the fidelity can well
reflect the different phases and reveal different phase
transitions.

Secondly, the FS is studied and the finite scaling analysis is
performed for the MMT and the MIT. The contour maps of FS can well
reflect the phase diagram. At the MMT, the critical exponents
$\alpha$($\beta$,$\nu$) for system sizes $F_{m=3l+1}$ and
$F_{m\neq 3l+1}$ are different, but the relation that
$\nu/\alpha\approx2$ is universal. At the MIT, the critical
exponents for all system sizes are same and the relation that
$\nu/\alpha\approx2$ is also tenable.

At last, the von Neumann entropy is studied. It is near one in the
metallic phase, while small in the insulating phase. There are
sharp changes at phase boundaries. There are difference in the
values of spectrum-average von Neumann entropy for the three phase
boundaries and the bicritical point. The contour maps of von
Neumann entropy is consistence with the phase diagram. All these
indicate that the different phases and phase transitions can be
completely distinguished by von Neumann entropy.

\begin{acknowledgments}
Project supported by the National Nature Science Foundation of
China (Grant Nos 90203009 and 10674072), by the Specialized
Research Fund for the Doctoral Program of Higher Education (Grant
No 20060319007), by the Excellent Young Teacher Program of the
Ministry of Education of China.
\end{acknowledgments}

\end{document}